\documentclass[8pt, preprint]{aastex}

\newcommand{\HI}{\ion{H}{1}}
\newcommand{\kms}{km s$^{-1}$}

\slugcomment{ApJS accepted}

\begin{document}

\title{Atomic and Molecular Gas in Colliding Galaxy Systems: I. The Data}
\author {Daisuke Iono\altaffilmark{1,2}, Min S. Yun\altaffilmark{1}, 
Paul T. P. Ho\altaffilmark{2}}

\altaffiltext{1}{Department of Astronomy, University of Massachusetts, 
Amherst, MA 01002} 
\altaffiltext{2}{Harvard-Smithsonian Center for Astrophysics, 60 Garden Street, Cambridge, MA 02138}

\begin{abstract}

We present \ion{H}{1} and CO~(1--0) interferometric observations of 10 
comparable-mass interacting systems obtained at the Very Large Array 
(VLA) and the Owens Valley Radio Observatory (OVRO) millimeter array. 
The primary intent
of this study is to investigate the response of cold gas during the 
early stages of collision of massive disk galaxies.  The sample sources 
are selected based on their luminosity (M$_B \leq -19$), 
projected separation (5-40 kpc),
and single dish CO~(1--0) content ($S_{CO} \geq 20$ Jy \kms). These 
selection criteria result in a sample that primarily consists of 
systems in the early stages of an interaction or a merger.   
Despite this sample selection, 50\% of the systems show long \HI\
tidal tails indicative of a tidal disruption in a prograde orbit.
In addition, all (4/4) of the infrared luminous pairs (LIRGs) 
in the sample show long \HI\ tails, suggesting that the presence
of a long \HI\ tail can be a possible signature of enhanced 
star formation activity in a collision of gas-rich galaxies.
More than half of the groups show a displacement of \HI\ peaks 
from the stellar disks.  
The CO~(1--0) distribution is generally
clumpy and widely distributed, unlike in most IR-selected late
stage mergers -- in fact, CO peaks are displaced from the stellar
nucleus in 20\% (4/18) of the galaxies with robust CO detection.  
\HI\ and CO~(1--0) Position Velocity Diagrams (PVDs) and rotation curves are 
also presented, and their comparison with the numerical simulation
analyzed in Paper~I show evidence for radial inflow and wide 
occurrences of nuclear molecular rings.
These results are further quantified by examining 
physical and structural parameters derived in comparison with 
isolated systems in the BIMA SONG sample in our forthcoming paper.

\end{abstract}


\section{Introduction}

The importance of galaxy collisions and mergers to the formation and 
evolution of galaxies has been clearly demonstrated by a variety of 
cosmological simulations \citep[e.g.][]{cole00}.  It is believed that a 
galaxy experiences a close encounter with another several times during 
its lifetime, from a merger involving two comparable mass galaxies 
(major merger) to  less catastrophic events involving a smaller companion 
or a satellite (minor merger) \citep[see][for a review]{struck99}.  
Cosmological simulations also show  that hierarchical galaxy formation 
is an ongoing process even in the present epoch as evidenced by the 
ubiquitous presence of systems in the local universe that show signatures 
of major and minor mergers \citep[e.g.][]{murali02}.  Observations of the 
early universe conducted using sensitive sub-mm detectors suggest that 
collisions and mergers played important roles to the formation of stars 
and galaxies in the early epochs \citep{blain02}.  Studying the colliding 
galaxy population in the local universe, therefore, is an important first 
step toward better understanding the process of galaxy formation and 
evolution and the star formation history in the universe.

Star formation requires an abundant supply of molecular gas. Studies of 
molecular gas (traced in CO) in the Milky Way has been done efficiently 
and extensively using single dish radio telescopes and interferometers.   
In extragalactic studies, extensive large scale surveys of molecular gas 
have been largely limited to single dish 
observations \citep[see ][]{young91,young95}.  A beam that subtends a 
large area of the sky allows an efficient determination of the total 
amount of the emission, and hence the total molecular gas mass,  
in a relatively short amount of time.  The resultant large 
database of hundreds of galaxies yields a statistically significant 
comparison of the physical properties in different Hubble types.   
A main shortcoming of these observations is the lack of spatial information 
resulting from the large beam sizes.    
Interferometric observations can significantly improve the resolution 
problem, but sparse sampling of the uv-space by the existing array
telescopes operating at millimeter wavelengths generally suffer
from poor surface brightness sensitivity and limited dynamic ranges
($\le 10-20$).  
In addition, an extensive survey to develop a sample size comparable to those 
achieved by single dish experiments would take a prohibitively long time.  
\citet{sakamoto99} were the first to conduct an extensive interferometric 
survey of CO~(1--0) emission in 20 nearby galaxies using 
the 6-element Nobeyama Array and the Owens Valley Radio Observatory (OVRO)\footnote{The Owens Valley Radio Observatory 
is operated by the California Institute of Technology.} millimeter
array. In a more recent survey using the 
Berkeley-Illinois-Maryland Association (BIMA) array \citep{helfer03}, 
44 nearby late type galaxies were imaged at $\sim6\arcsec$  
resolution. Thus a growing database of high resolution CO~(1--0) observations
of nearby spiral galaxies is now available.

Single dish telescope surveys dedicated exclusively to measuring 
CO~(1--0) emission in colliding disk systems have been carried out previously 
\citep{zhu99, gao99, georgakakis00, yao03}, but interferometric
surveys of CO emission with good spatial information are limited in 
number and in size.  The \ion{H}{1} 
emission in colliding galaxies has been studied both using single dish and 
interferometers, and the characteristics of the \ion{H}{1} tidal tails has 
been used as an important diagnostics to trace the history of the interaction 
and star formation activity \citep[e.g.][]{hibbard96}.  The important next 
step, therefore, is to study the spatial distribution and the kinematics of 
dense and diffuse molecular ISM in colliding systems with the 
high resolution afforded by interferometers.    

The response of the gas in a simulated disk-disk collision was investigated 
in detail by \citet[][Paper~I hereafter]{iono04}.  It was found that stars 
respond to the tidal interaction by forming both transient arms and long 
lived $m=2$ bars, but the gas response is more transient, flowing directly 
toward the central regions within about $10^8$ years after the initial 
collision.  Comparing the predicted inflow timescale 
($10^8$ years) with the total merger timescale 
($5 - 10 \times 10^8$ years) suggests that 10 - 20\% of randomly 
selected interacting/merging systems in the local universe have a 
possibility to exhibit observable signature of radial inflow.  The 
evolution of the structural parameters such as the asymmetry ($A$), 
the concentration ($C$) and the compactness ($K$) parameters were 
investigated, and the possible use of the $K$ parameter and the molecular 
fraction ($M_{H_2}/M_{gas}$) to infer the merger chronology was suggested.   
It was shown that distinct emission features in the 
forbidden velocity quadrants  of the position velocity diagram (PVD) 
identifies non-circular gas kinematics driven by the perturbation of the 
non-axisymmetric structure.  These diagnostics tools developed using 
numerical simulation can be applied directly to observational data, and 
whether the same behavior is seen in observations is an important test 
concerning the validity of the physics implemented in the numerical 
simulations.  To this end, this paper describes  the observational data 
of 10 colliding systems in \ion{H}{1} and CO~(1--0) emission obtained 
using the Very Large Array (VLA)\footnote{The National Radio Astronomy 
Observatory is a facility of the National Science Foundation operated 
under cooperative agreement by Associated Universities, Inc.} and Owens 
Valley Radio Observatory millimeter 
array.  Detailed statistical and comparative  analysis of this data with 
a sample of nearby isolated systems can be found in our forthcoming paper 
(Paper~III).  

The organization of this paper is as follows.
First, the sample selection is described in \S2, followed by a 
description of the data reduction and calibration techniques in \S3.  
Main results are presented and discussed in \S4, \S5, \S6 and \S7.   
Qualitative and quantitative descriptions of the individual sources 
are presented in \S8.  A short summary is given in \S9.

\section{Sample Selection}

A significant amount of information pertaining to the interaction history 
can be obtained by examining the optical morphology alone. \citet{bushouse86} 
compiled a set of $\sim 100$ strongly interacting systems that involve two 
or more progenitor galaxies by visually inspecting the Uppsala General 
Catalogue (UGC). He concluded that these systems show a systematically 
higher level of H$\alpha$ luminosity and equivalent width than those found 
in isolated systems, but also found that about 30\% of the interacting 
galaxies show weak or no optical emission lines much like the spectra of 
elliptical galaxies.  A sub-set of 80 systems were further investigated by 
\citet{zhu99} in CO~(1-0) using both the NRAO 12~m and the IRAM 30~m single 
dish radio telescopes.  The interacting systems investigated in this study 
have been selected from the sample of \citet{zhu99} according to the 
following criteria:
\begin{enumerate}

\item \textbf{Major Mergers:} systems must include two nearly equal mass large spirals with $M_B \leq -19$,

\item \textbf{Projected Separation:} interacting pairs covering a wide range of separation, from 5 to 40 kpc, are chosen,

\item \textbf{Proximity:} sufficiently close ($\lesssim 100$ Mpc) to be 
resolved with the OVRO resolution ($\sim 5''= 2.4$~kpc at D=100~Mpc),

\item \textbf{CO emission:} required $S_{CO} \geq 20$ Jy km s$^{-1}$ 
to ensure clear detection by the interferometer.

\end{enumerate}

Out of the 80 pairs in \citet{zhu99}, eight pairs meet this selection 
criteria.  Five sources (VV~48, VV~247, VV~769, 
VV~242, VV~244) were observed in CO~(1--0)
using the OVRO in Spring 2002, two sources (VV~246, VV~254) 
were observed during the 1997--1998 season, 
and the final source (VV~253) was observed in Spring 1995 and the data were 
retrieved from the OVRO archive.   
In addition to the eight sources selected from the \citet{zhu99} sample, 
three pairs that satisfy the same selection criteria (VV~219, VV~55, VV~254)  
were observed at the OVRO array.  
VV~219 was observed in Spring 2002, while both VV~55 and VV~254 were observed
during the 1997--1998 season.  Observation of VV~246 (NGC~3395/6) 
resulted in a non-detection by the interferometer,
despite the reported flux of 58~Jy~km~s$^{-1}$ (NGC~3395) 
and 39~Jy~km~s$^{-1}$ (NGC~3396)
obtained using the NRAO~12~m telescope \citep{zhu99}.
The non-detection result of VV~246 is not presented here.
This will comprise an interferometric CO~(1--0) sample of a total of 
10 pairs (or 20 galaxies)(see Table~\ref{galaxy_sample}) 
-- a large enough of a sample to allow the full examination
of the large parameter space involved.
Five of these sources (VV~55, VV~253, VV~242, VV~219, VV~244) 
were also observed in \ion{H}{1} emission using 
the VLA in Winter 2002.   The \ion{H}{1} data for three 
sources (VV~48, VV~247, VV~731) were obtained from the VLA archive, 
while the maps of VV~254 and VV~769 were kindly supplied by 
 J. Condon.


To place the characteristics of the sample sources in a broader context, 
a first order assessment of the merger chronology is performed by sorting 
the sample systems according to their projected nuclear separation 
(Table~\ref{galaxy_sample} and Figure~\ref{dss_all}).  This 
classification scheme is similar to the method adopted by 
\citet[][i.e. the Toomre Sequence]{toomre77}  where eleven colliding/merging 
galaxy pairs from various stages of the interaction were selected from the NGC 
catalog, and the optical morphology of the pairs were used to roughly 
define the interaction sequence. 
Classifying the interacting galaxies in this 
way can introduce significant uncertainties because the sample sources 
are predominantly early to intermediate stage systems, whereas the Toomre 
sequence covers a much broader range of interacting systems from the 
early stage (``the Antennae (NGC~4038/9)'') to the late stage merger
(``Atoms for Peace (NGC~7252)'').   This ordering will be 
revised in Paper~III from a  different perspective using a more 
comprehensive and quantitative set of analysis tools derived from the global 
properties in optical, CO~(1--0) and \ion{H}{1} emission.

In order to investigate the environment in which each galaxy pair 
resides, the projected 2~Mpc radius of each pair was searched for the
presence of neighboring 
galaxies using the NASA/IPAC Extragalactic Database 
(NED)\footnote{This research has made use of the NASA/IPAC Extragalactic 
Database (NED) which is operated by the Jet Propulsion Laboratory, 
California Institute of Technology, under contract with the National 
Aeronautics and Space Administration.}.  Constraining the velocity range
according to the Hubble Law will only include sources 
in a narrow velocity space (i.e. $\Delta v = 150$ km s$^{-1}$ for 2~Mpc 
using $H_0 = 75$ km s$^{-1}$ Mpc$^{-1}$),
and it will likely preclude galaxies with high peculiar velocities 
with respect to the Hubble flow.  
To be conservative, therefore, the adopted velocity 
range includes galaxies within $\Delta v = 500$ km s$^{-1}$
of the systemic velocity.
The results are presented in Table~\ref{environment}.  
The NED only includes galaxies that exist in 
previous survey catalogs and may not have the faint dwarf companions that 
past instruments were not capable of detecting due to sensitivity or 
coverage limitations.  For example, the all sky 2MASS catalog 
\citep{jarrett03} and the SDSS \citep{abazajian03} 
has a sensitivity limit of $K_s = 13.5$ and $r = 22$, respectively.
Excluding such faint undetected dwarf galaxies 
which are predicted to be abundant, half of the program sources (VV~55, 
VV~48, VV~253, VV~219, VV~244) appear
to reside in a high density environment as the large number of
neighboring sources indicates.  VV~219 is an interacting system in 
the Virgo Cluster and therefore has an exceptionally large 
number of neighbors.  The rest of the systems (VV~254, VV~247, VV~769, 
VV~242, VV~731) have about a factor of ten smaller number ($N \simeq 0 - 6$) 
of neighboring sources, which suggests that the evolution is determined 
mostly from the companion galaxy in the respective galaxy systems.

\begin{figure}
  \caption{DSS2 images of the sample sources arranged by decreasing projected 
nuclear separation.  
The horizontal bar drawn in each panel shows the physical scale of 10~kpc, 
with the angular size labeled in arc-seconds.
The abscissa and ordinate
are aligned with east-west and north-south directions respectively.
See \citet[][Fig.~3]{keel03}
for a high resolution \textit{HST} image of VV~247.  The high quality 
\textit{HST} image clearly shows the detailed structure of the nuclear regions
as well as the ubiquitous dust lanes outlining the tidal tails. 
The Second Palomar Observatory Sky 
Survey (POSS-II) was made by the California Institute of Technology with 
funds from the National Science Foundation, the National Geographic Society, 
the Sloan Foundation, the Samuel Oschin Foundation, and the Eastman Kodak 
Corporation.}
  \label{dss_all}
\end{figure}

\begin{deluxetable}{lcccccccccc}
\tabletypesize{\scriptsize}
\tablewidth{0pt}
\tablecaption{Interacting Galaxy Sample\label{galaxy_sample}}
\tablehead{
\colhead{Source}& \colhead{R.A.\tablenotemark{a}} &  \colhead{Decl.} & \colhead{
Class.\tablenotemark{b}}   & \colhead{M$_B$\tablenotemark{c}} & \colhead{z ($D$)
\tablenotemark{d}} & \colhead{sep.\tablenotemark{e}}  & \colhead{$L_{(F)IR}$\tablenotemark{f}} & \colhead{Ratio of L$_B$\tablenotemark{g}}\\
& \colhead{(J2000)} &  \colhead{(J2000)} &  &  & \colhead{(Mpc)} & \colhead{(kpc)}  & \colhead{($10^{10} \rm L_{\odot}$)} &  (L$_1$/L$_2$)
} 
\startdata
VV~55            &             &            &     &     & 0.023 (90) & 38   & 28.3 (15.1) & 1.00\\
~~~NGC 5257  & $13^h 39^m 52.9^s$ & $+00^d 50^m 24.5^s$  & HII & -21.8\\
~~~NGC 5258  & $13^h 39^m 57.7^s$ & $+00^d 49^m 51.5^s$  & HII/L & -21.7\\
VV~48            &             &            &     &     & 0.012 (42) & 26   & 6.1 (3.2) & 1.07\\
~~~NGC 5394  & $13^h 58^m 33.7^s$ & $+37^d 27^m 12.5^s$  & \nodata & -19.8\\
~~~NGC 5395  & $13^h 58^m 37.9^s$ & $+37^d 25^m 28.5^s$  & \nodata & -21.1\\
VV~254  &             &            &     &     & 0.015 (55) & 20   & 7.0 (3.6) & 1.03\\ 
~~~UGC 12914 & $00^h 01^m 38.3^s$ & $+23^d 29^m 01.2^s$ & L & -21.2\\
~~~UGC 12915 & $00^h 01^m 41.9^s$ & $+23^d 29^m 45.2^s$ & L & -20.6\\
VV~253           &             &            &     &     & 0.033 (131)& 18   &  32.3 (17.9) & \nodata\\
~~~NGC 5331N & $13^h 52^m 16.4^s$ & $+02^d 06^m 31.5^s$  & HII & -21.6\\
~~~NGC 5331S & $13^h 52^m 16.2^s$ & $+02^d 06^m 04.5^s$  & \nodata  & (total)\\
VV~247           &             &            &     &     & 0.021 (81) & 17   & 14.2 (7.6) & 1.11\\
~~~NGC 6621  & $18^h 12^m 55.3^s$ & $+68^d 21^m 48.5^s$ & HII  & -20.6\\
~~~NGC 6622  & $18^h 12^m 59.6^s$ & $+68^d 21^m 14.5^s$  &  \nodata  & -18.5\\
VV~769 &             &            &     &     & 0.018 (67) & 17   & 4.5 (2.4) & 1.02 \\
~~~UGC 813   & $01^h 16^m 16.5^s$ & $+46^d 44^m 24.8^s$ &\nodata & -20.4 \\
~~~UGC 816   & $01^h 16^m 20.5^s$ & $+46^d 44^m 52.8^s$ &\nodata & -20.9 \\
VV~242           &             &            &     &     & 0.016 (57) & 13 & 7.1 (3.9) & 1.00\\
~~~UGC 11984  & $22^h 19^m 27.8^s$ & $+29^d 23^m 44.9^s$  & \nodata & -20.5\\
~~~UGC 11985  & $22^h 19^m 30.3^s$ & $+29^d 23^m 16.9^s$  & \nodata & -20.5\\
VV~219           &             &            &     &     & 0.004 (16)\tablenotemark{h} & 11 & 2.2 (1.1) & 1.03 \\
~~~NGC 4567  & $12^h 36^m 32.7^s$ & $+11^d 15^m 29.0^s$  &\nodata & -20.5\\
~~~NGC 4568  & $12^h 36^m 34.3^s$ & $+11^d 14^m 20.0^s$  & \nodata& -21.1\\
VV~731           &             &            &     &     & 0.025 (98) & 7 & 21.5 (13.6) & \nodata \\
~~~NGC 7592W & $23^h 18^m 21.8^s$ & $-04^d 24^m 57.1^s$  & HII & -21.1\\
~~~NGC 7592E & $23^h 18^m 22.6^s$ & $-04^d 24^m 58.1^s$  & \nodata& (total)\\
VV~244           &             &            &     &     & 0.007 (24) & 5 & 2.2 (1.1) & 1.01\\
~~~NGC 5953  & $15^h 34^m 32.4^s$ & $+15^d 11^m 37.8^s$  & L/S2  & -19.2  \\
~~~NGC 5954  & $15^h 34^m 35.0^s$ & $+15^d 12^m 00.8^s$  & S2    & -19.1 \\
\enddata
\tablenotetext{a}{The source coordinates are found from the peak of the 2MASS K-band images.  The resolution of the images is $2-3''$. Atlas Image obtained as part of the Two Micron All Sky Survey (2MASS), a joint project of the University of Massachusetts and the Infrared Processing and Analysis Center/California Institute of Technology, funded by the National Aeronautics and Space Administration and the National Science Foundation.}
\tablenotetext{b}{Spectral classification obtained from NED, where different letters correspond to; LINER (L), star forming (HII) and Seyfert~2 (S2). }  
\tablenotetext{c}{From RC3 when available, else from \citet{georgakakis00}(NGC~5331), \citet{bushouse87}(NGC~6621/2), \citet{zhu99}(UGC~11984/5) and \citet{soifer87}(NGC~7592).}  
\tablenotetext{d}{The redshift information is retrieved from NED.  The luminosity  distance is derived using the $\Lambda$CDM cosmology ($H_0 = 75$ km s$^{-1}$ Mpc$^{-1}$, $\Omega_M = 0.3$, $\Omega_\Lambda = 0.7$)}  
\tablenotetext{e}{Projected physical separation of the pair.}  
\tablenotetext{f}{The infrared and far infrared luminosities are derived from the IRAS 12, 25, 60 and $100\micron$ flux in the \textit{IRAS} Revised Bright Galaxy Sample \citep{sanders03} for all of the systems except for VV~769 where the values from the Faint Source Catalogue \citep{moshir90} is used. $L_{IR} = 4 \pi D^2_L F_{IR}$, where $F_{IR} = 1.26 \times 10^{-14} [13.48~f_{12} + 5.16~f_{25} + 2.58~f_{60} + f_{100}$] \citep{sanders96}.  The FIR luminosities are shown in () \citep{helou88}.}  
\tablenotetext{g}{Ratio of the absolute magnitudes in the pair should reflect the relative masses (assuming a constant mass-to-light ratio).  Values for VV~253 and VV~731 are not available because B-band magnitudes are available only for the whole systems.}  
\tablenotetext{h}{The redshift of VV~219 in NED is given as $z=0.0075$ (or $D=28$~Mpc), but the redshift of the giant elliptical galaxy at the center of the Virgo cluster M87 is given as $z=0.0044$ (or $D=16$~Mpc).  The discrepancy may be due to the high peculiar velocity of VV~219 in the cluster potential.  Because VV~219 and M87 are located well within a projected distance of 1~Mpc, we adopt $z=0.0044$ (or $D=16$~Mpc) for the redshift of VV~219.}
\end{deluxetable}

\begin{deluxetable}{lcclcccccc}
\tabletypesize{\scriptsize}
\tablewidth{0pt}
\tablecolumns{6}
\tablecaption{Galaxies Within 2 Mpc Volume Around the Sample Sources\label{environment}}
\tablehead{
\colhead{Source}& \colhead{$cz$\tablenotemark{a}} & \colhead{No. of sources} & \colhead{Nearby Source\tablenotemark{b}} & \colhead{Projected Distance} & \colhead{$v$} \\
 &\colhead{(km s$^{-1}$)}& & & \colhead{(kpc)} & \colhead{(km s$^{-1}$)} 
} 
\startdata
VV~55& $6777 \pm 500$ & 12 & USGC U556 & 233 & 6785 \\
&&&IC~904 & 709 & 6752 \\
&&&WBL~460 & 715 & 6805 \\
&&&CGCG~017-046 & 757 & 6754 \\
&&&2dFGRS~N335Z175 & 825 & 6727 \\
VV~48& $3481 \pm 500$ & 23 & USGC U585 & 148 & 3286 \\
&&&[FK2002] 207 & 159 & 3270 \\
&&&NGC 5380 & 263 & 3173 \\
&&&NGC 5395 & 303 & 3290 \\
&&&NGC 5378 & 354 & 3042 \\
VV~254 & $4353 \pm 500$ & 6 & KUG 2357+228 & 515 & 4461 \\
&&&UGC 24 & 1035 & 4442 \\
&&&KUG 0003+235 & 1085 & 4574 \\
&&&UGC 11 & 1379 & 4447 \\
&&&NGC 9 & 1632 & 4528 \\
VV~253 & $9906 \pm 500$ & 17 & [WZX98] 13496+0221C & 0 & 9833 \\
&&&CGCG 017-081 & 19 & 9766 \\
&&&WBL 471 & 50 & 9926 \\
&&&USGC U572 & 240 & 9787 \\
&&&2MASX J13513864+0209279 & 324 & 9650 \\
VV~247 & $6329 \pm 500$  & 1 & UGC~11183 & 544 & 6164\\
VV~769 & $5266 \pm 500$  & 1 & CGCG 551-011 & 43 & 5373\\
VV~242 & $4531 \pm 500$  & 1 & CGCG 494-017 & 807 & 4583\\
VV~219 & $2265 \pm 500$ & 58 & IC 3509 & 478 & 2000 \\
&&&NGC~4607 & 636 & 2257\\
&&&IC~3562 & 654 & 2051\\
&&&NGC~4569 & 666 & 1870\\
&&&VCC~1919 & 770 & 1869\\
VV~731 & $7327 \pm 500$ & 0 & \nodata & \nodata & \nodata \\
VV~244 & $1962 \pm 500$ & 20 & UGC~9902 & 25 & 1696\\
&&& KTG 62 & 54 & 1906\\
&&& NGC 5951 & 114 & 1780\\
&&& NGC 5962 & 321 & 2040\\
&&& UGC~9951 & 491 & 2004\\
\enddata
\tablenotetext{a}{The redshifts listed in Table~\ref{galaxy_sample} 
are converted to velocity, and the search radius includes galaxies 
within $\pm 500$~km~s$^{-1}$ of this velocity}  
\tablenotetext{b}{The source database includes the standard catalogs
such as the NGC and UGC, as well as the more recent deep surveys such as
the 2MASS \citep{jarrett03} and SDSS \citep{abazajian03}, 
which contains galaxies brighter than $K_s = 13.5$ and $r = 22$, respectively.
The closest five sources are listed here if the number of sources 
is greater than five.}
\end{deluxetable}

\section{Observations and Data Reduction}

\subsection{\ion{H}{1}}

The \ion{H}{1} observations for five of the sources (VV~55, VV~253, 
VV~242, VV~219, VV~244) were carried out at the VLA in Winter 2002 using 
the C configuration.  The correlator was configured to use 2 IFs with 
3.125 MHz total bandwidth and 48.8 kHz (10.5 km s$^{-1}$) frequency 
resolution after on-line Hanning smoothing.   The on-source integration 
time for each source was $\sim 5$ hours.  A nearby quasar was 
observed every 30-40 minutes to track the gain variation.
The flux and passband calibrations were performed by observing
the bright standard flux calibrators 3C~48 (16.5~Jy at 1.4~GHz) 
or 3C~286 (15.0~Jy at 1.4~GHz).  The calibration and imaging were 
carried out using the NRAO software system AIPS.  Natural
weighting (ROBUST = 5) of the visibility data was adopted to maximize 
the sensitivity for the imaging.  Key details of
the observations are summarized in Table~\ref{vla_properties}.

\subsection{CO~(1--0)}

The CO~(1--0) observations for six of the sources (VV~48, VV~247, VV~769, 
VV~242, VV~219, VV~244) were carried out using the  OVRO interferometer array 
during the Spring 2002 season.   Two antennas were configured in the 
north-south direction with a 50 m baseline length while four antennas with 
baseline lengths between 15 and 115 m were assigned in the east-west 
direction (the L configuration).  The digital correlator was configured 
to cover 480 MHz in 4 modules, each divided into 120 channels giving a 
velocity resolution of $\sim$ 10.5 km s$^{-1}$.  
One or more bright quasars such as 3C~84, 3C~273, or 3C~454.3 were
observed in each track for the passband calibration. 
The absolute flux scale is based on the observations of Uranus.
A nearby quasar  
selected for each target source was observed every 20 minutes to track the 
short term gain variation, and amplitude and phase fittings were performed 
using  baseline based calibration.   The data were reduced using the OVRO 
data reduction program MMA \citep{Scoville93}. 
The calibrated data were imaged and deconvolved using the
program DIFMAP \citep{shep94}. Two narrow band channels were 
averaged to produce a natural weighted CLEANed data cube at a final velocity 
resolution of 21 km s$^{-1}$.   For sources with low S/N,  a Gaussian taper 
was applied in the uv-space to smooth the data spatially.  
For sources with high S/N, uniform weighting 
was used to improve the angular resolution.  Key details of
the observations are summarized in Table~\ref{ovro_properties}.

\begin{deluxetable}{lcccccc}
\tablewidth{0pt}
\tablecaption{VLA Observational Properties\label{vla_properties}}
\tablehead{
\colhead{Source} &  \colhead{Date} & \colhead{Config.} & \colhead{$\sigma_{\rm RMS}$\tablenotemark{a}} & \colhead{Beam\tablenotemark{b}} & \colhead{Beam} & \colhead{Calibrator}\\
 & & & \colhead{(mJy)} & \colhead{($''$)} & \colhead{(kpc)} & 
} 
\startdata
VV~55 &Winter 2002 &C & 0.18  & $22.2 \times 17.6$ & $9.8 \times 7.8$ & J1354$-$021\\
VV~48 & Summer 1993 &C,D& 0.79 & $17.8 \times 16.3$ & $4.0 \times 3.6$ & J1504+377\\
VV~254  & Fall 1990 & C & \nodata & $18.0 \times 18.0$ & $5.1 \times 5.1$ & J2330+110 \\
VV~253  & Winter 2002 & C & 0.22 & $22.0 \times 17.1$ & $14.2 \times 11.0$ & J1354$-$021\\
VV~247   & Winter 1998 & CnB & 0.57 & $16.2 \times 13.0$ & $6.5 \times 5.2$ & J2236+284\\
VV~769   & Spring 2001 & B,C & 0.23 & $16.3 \times 16.3$ & $5.7 \times 5.7$ & J0114+483 \\
VV~242 &Winter 2002 &C& 0.18 & $17.9 \times 17.6$ & $5.5 \times 5.4$ & J1252+119\\
VV~219 &Winter 2002 &C& 0.86  & $19.9 \times 14.0$ & $1.6 \times 1.1$ & J1252+119\\
VV~731 &Summer 1987 &D & 1.95  & $128.6 \times 82.6$ & $35.5 \times 22.8$ & J2253+161\\
VV~244 &Winter 2002 &C& 0.29  & $19.8 \times 17.9$ & $2.3 \times 2.1$ & J1520+202\\
\enddata
\tablenotetext{a}{The RMS noise per 11 km s$^{-1}$ channel, except for VV~731 where 21 km s$^{-1}$ velocity resolution was used.  Data for VV~254 and VV~769 were provided by J. Condon \citep{cond93, cond02}, and the noise properties for VV~254 is unknown because the map was clipped at a threshold value.} 
\tablenotetext{b}{The synthesized beam size recovered using natural weighting (ROBUST = 5).}
\end{deluxetable}

\begin{deluxetable}{lcccc}
\tablewidth{0pt}
\tabletypesize{\small}
\tablecaption{OVRO Observational Properties\label{ovro_properties}}
\tablehead{
\colhead{Source}  & \colhead{$\sigma_{\rm RMS}$ \tablenotemark{a}} & \colhead{Beam\tablenotemark{b}} & \colhead{Beam} & \colhead{Calibrator
}\\
 &  \colhead{(mJy)} & \colhead{($''$)} & \colhead{(kpc)} &  
} 
\startdata
VV~55   & 16 & $6.2 \times 3.7$& $2.7 \times 1.6$ &  J1256$-$057\\
VV~48    & 16 & $5.5 \times 4.3$& $1.2 \times 1.0$  & J1153+495\\
VV~254  & 15 & $7.2 \times 5.1$ & $2.0 \times 1.4$  & J2253+161\\
        &(20)& ($4.3 \times 3.4$) &  ($1.2 \times 1.0$)  &  \nodata \\
VV~253  & 11 & $5.4 \times 4.4$& $3.5 \times 2.8$  &  J1256$-$057\\
VV~247  & 24 & $6.6 \times 5.6$& $2.7 \times 2.3$  & J1800+784 \\
VV~769  & 16 & $5.1 \times 4.1$& $1.8 \times 1.4$  & J0136+478 \\
VV~242  & 11 & $4.4 \times 3.1$& $1.3 \times 0.9$ & J1549+026\\
VV~219  & 12 & $4.0 \times 3.5$& $0.3 \times 0.3$  & J1549+026 \\
VV~731  & 10 & $4.5 \times 3.6$& $2.2 \times 1.7$  & J2246$-$121\\
VV~244  & 15 & $4.4 \times 3.6$& $0.6 \times 0.5$  & J1549+026\\
\enddata
\tablenotetext{a}{The RMS noise per 21 km s$^{-1}$ channel} 
\tablenotetext{b}{The synthesized beam size recovered using natural weighting, except for VV~254 where uniform weighting was also adopted in order to increase the resolution (shown in ()).}
\end{deluxetable}

\section{Derived Properties of Atomic and Molecular Gas}

The derived quantities from the \ion{H}{1} and CO~(1--0)  observations 
are summarized in Table~\ref{derived_HI} and \ref{derived_CO} respectively. 
The properties of \ion{H}{1} were derived for the whole system, rather than 
trying to separate the \ion{H}{1} into two galaxies.  Table~\ref{derived_HI} 
has the following format.\\

\noindent Col. (1)  Name of the pair from the Vorontsov-Velyaminov catalog \citep{vv59, vv77}\\

\noindent Col. (2)  Recovered integrated flux derived from the moment zero map.  The AIPS task MOMNT was used to generate the  moment zero map.\\

\noindent Col. (3)  Total atomic gas mass calculated assuming an optically thin emission, $M_{HI} (M_\odot) = 2.36 \times 10^{5} D^2 \int S_\nu dv$
where $D$ is distance in Mpc and $\int S_\nu dv$ in Jy \kms.  The integrated flux ($\int S_\nu dv$) given in Col. (2) is used.  \\

\noindent Col. (4) Peak brightness temperature.\\

\noindent Col. (5) Peak column density.\\

\noindent Col. (6) Total line width in Full Width Zero Intensity (FWZI).\\

Table~\ref{derived_CO}  has the following format:\\

\noindent Col. (1)  Name of the pair from the Vorontsov-Velyaminov catalog \citep{vv59, vv77} and the UGC/NGC number for each individual galaxy.\\

\noindent Col. (2)  Recovered integrated flux derived from the moment zero map.
The AIPS task MOMNT was used to generate the moment zero map.  Single dish measurements from \citet{zhu99} are listed for comparison in () when available.\\

\noindent Col. (3)  Derived total molecular gas mass from the integrated flux in Col. (2) when available. The Galactic CO to H$_2$ conversion ($M_{H_2} (M_{\odot}) = 1.18 \times 10^4 D^2 \int S_{CO(1-0)} dv$, where $D$ is distance in Mpc and $\int S_{CO(1-0)} dv$ is the integrated interferometer flux in Jy \kms) was used. \\

\noindent Col. (4)  Dynamical mass derived from the maximum radial extent of the CO~(1--0) emission and the observed velocity maxima.  Inclination correction is not applied. \\

\noindent Col. (5)  Peak brightness temperature.\\

\noindent Col. (6)  Peak H$_2$ column density.\\

\noindent Col. (7) Average surface density given by $M_{H_2}/\pi R^2 ln(2)$, where $R$ is the FWHM of the deconvolved size of the emission.\\

\noindent Col. (8) Systemic velocity derived from the median velocity of the CO~(1--0) spectrum.\\

\noindent Col. (9) Estimated total line width in Full Width Zero Intensity (FWZI).\\

\noindent Col. (10) Ratio of molecular mass to dynamical mass.  Note that M$_{dyn}$ is not corrected for the inclination.\\

\noindent Col. (11) Ratio of molecular gas to total gas mass.  The total gas mass is the sum of atomic (from Table~\ref{derived_HI}) and molecular gas (from Col. (3)) for the whole system.\\

\noindent Col. (12) Star formation efficiency ($L_{IR}/M_{H_2}$).  The molecular gas mass derived from the interferometer in Col. (3) is used.\\

\begin{deluxetable}{lrrrrr}
\tabletypesize{\small}
\tablewidth{0pt}
\tablecaption{Derived \ion{H}{1} Properties\label{derived_HI}}
\tablehead{
\colhead{Source}&\colhead{$\rm S_{\nu} dv$} & \colhead{log$\rm M_{HI}$}&\colhead{$\Delta \rm T_{peak}$} &\colhead{log$\rm N_{peak}$} & \colhead{$\Delta \rm V_{ZI}$} \\
 & (Jy km s$^{-1}$) & (M$_{\odot}$) & (K) & (cm$^{-2}$) & (km s$^{-1}$) 
} 
\startdata
VV~55                    & 17.6  & 10.5 & 7.9 & 21.4  & 475\\
VV~48                    & 13.8  & 9.8 & 20.3 & 21.6 & 620 \\
VV~254\tablenotemark{a}  & 14.0  & 10.1 & 12.4 & 21.5 & 721 \\
VV~253                   & 4.8   & 10.3 & 29.0 & 21.3  & 573 \\
VV~247                   & 2.6  & 9.6 & 10.5 & 21.3 & 518 \\
VV~769\tablenotemark{b}  & 9.4  & 10.1 & 9.4 & 21.6 & 854 \\
VV~242                   & 11.5  & 10.1 & 21.9 & 21.7 & 447 \\
VV~219                   & 21.1  & 9.1 & 29.8 & 21.5 & 335 \\
VV~731                   & 0.8  & 9.2 & 2.6   & 19.8 & 87  \\
VV~244                   & 7.9  & 9.0  & 71.1 & 21.4 & 365 \\
\enddata
\tablenotetext{a}{\citet{cond93}}
\tablenotetext{b}{\citet{cond02}}
\end{deluxetable}

\begin{deluxetable}{lrrrrrrrrrrrr}
\tabletypesize{\scriptsize}
\rotate
\tablewidth{0pt}
\tablecaption{Derived CO(1--0) Properties\label{derived_CO}}
\tablehead{
\colhead{Source} & \colhead{$\rm S_{\nu} dv$\tablenotemark{a}} & \colhead{log$\rm M_{H_2}$} &\colhead{log$\rm M_{dyn}$} &\colhead{$\Delta \rm T_{peak}$} &  \colhead{log$\rm N_{peak}$} &\colhead{$\Sigma$} & \colhead{$v_{\rm sys}$} & \colhead{$\Delta \rm V_{ZI}$} &  \colhead{$\rm \frac{M_{H_2}}{M_{dyn}}$} &  \colhead{$\rm \frac{M_{H_2}}{M_{gas}}$\tablenotemark{b}} & \colhead{$\rm \frac{L_{IR}}{M_{H_2}}$}\\
 & (Jy km s$^{-1}$) & (M$_{\odot}$) & (M$_{\odot}$) & (K) & (cm$^{-2} (\rm M_{\odot}$ $\rm pc^{-2})$) & ($\rm M_{\odot}$ $\rm pc^{-2}$) & (km s$^{-1}$) & (km s$^{-1}$) & & & ($\rm L_\odot$ $M_\odot^{-1}$)
} 
\startdata
VV~55        & & & & & & & & & &0.52 &7.8\\
~~~NGC~5257   & 137  & 10.1 & 10.5 & 0.6 & 22.5 (471) & 108 & 6764 & 545 & 0.46  \\
~~~NGC~5258   & 250   & 10.4 & 11.0  & 1.1 & 22.6 (622) & 141 & 6775 & 479 & 0.22 \\
VV~48        & & & & & & & &&  & 0.49 &7.6\\
~~~NGC~5394   & 148 (201) & 9.5 & 9.3 & 2.1 & 22.8 (1021) & 189 & 3444 & 150 & 1.74 \\
~~~NGC~5395   & 171 (548) & 9.6 & 11.2 & 0.3 & 22.0 (168) & 68   & 3508 & 533 & 0.02 \\
VV~254        & & & & & & & &&  & 0.60 & 3.6\\
~~~UGC~12914  & 163 (420) & 9.8 & 11.0 & 0.7 & 22.3 (296)   & 71  & 4330 & 600 & 0.06 \\
~~~UGC~12915  & 334 (381) & 10.1  & 10.8 & 1.4 & 22.6 (681) & 106 & 4502 & 645 & 0.18 \\
VV~253        & & & & & & & &&  & 0.63 & 6.4\\
~~~NGC~5331N  & 17 (66) & 9.5  & \nodata  &  0.3 &22.2 (225) & 86 & 9860 & 440 & \nodata \\
~~~NGC~5331S & 150 (134) & 10.5 & 11.4 & 0.7 & 23.0 (1401) & 328 & 9920 & 690 & 0.13 \\
VV~247        & & & & & & & &&  & 0.9 & 7.1\\
~~~NGC~6621   & 369 (219) & 10.5 & 10.6 & 0.7 & 22.7 (815) & 150 & 6164 & 455 &0.72  \\ 
~~~NGC~6622   & \nodata (34) & \nodata & \nodata &\nodata & \nodata & \nodata& \nodata&\nodata & \nodata\\
VV~769        & & & & & & &&  & & 0.28 & 8.7\\
~~~UGC~813    &  27 (43)  & 9.2 & 10.1 & 0.4 & 22.2 (242) & 83 & 5159 & 345&0.11  \\
~~~UGC~816    & 60 (161) & 9.5  & 10.7 & 0.6 & 22.2 (256) & 73  & 5320 & 367&0.07 \\
VV~242        & & & & & & &&  & & 0.38 & 8.9\\
~~~UGC~11984  & 172 (180) & 9.8 & 10.7 & 0.7 & 23.0 (1654) & 255 & 4577 & 451&0.14  \\ 
~~~UGC~11985  & \nodata (32) & \nodata &\nodata & \nodata & \nodata & \nodata& \nodata&\nodata & \nodata\\
VV~219        & & & & & & &&  & & 0.64 & 2.7\\
~~~NGC~4567   & 121  & 8.6 & 9.5 & 1.2 & 22.2 (213) & 189  & 2285 & 169&0.12  \\
~~~NGC~4568   & 642  & 9.3 & 10.6 & 2.7 & 22.8 (960) & 426 & 2232 & 317&0.05  \\
VV~731        & & & & & & &&  & & 0.89 & 15.8\\
~~~NGC~7592W  & 59  & 9.8 & 9.9 & 0.8 & 22.7 (714)& 170 & 7353 & 416 &0.78  \\
~~~NGC~7592E  & 59  & 9.8 & 10.5 & 0.8 & 22.6 (610)& 91  & 7342 & 416&0.24  \\
VV~244        & & & & & & & & & & 0.68 & 8.0\\
~~~NGC~5953   &233 (365) & 9.2 & 9.7 & 2.1 & 22.7 (706) & 201 & 1968 & 254&0.35  \\
~~~NGC~5954   & 108 (73) & 8.9 & 9.5 & 1.2 & 22.5 (512) & 101 & 1984 & 253&0.21  \\
\enddata
\tablenotetext{a}{Single dish measurements from \citet{zhu99} are shown for comparison in () when available.  All of the single dish data were obtained at the NRAO~12~m telescope, except for VV~253 where the IRAM~30~m telescope was used.  In addition, a fully sampled map of VV~247 using the IRAM~30~m is presented in \citet{zhu99}, and this flux measurement is consistent with the sum of the fluxes of the pair obtained using the NRAO~12~m (see text).}
\tablenotetext{b}{The molecular gas mass fraction.  M$_{\rm gas}$ = M$_{\rm H_2}$ + M$_{\rm HI}$.}
\end{deluxetable}

\subsection{Flux Recovery and Derivation of Molecular Gas Mass}

The main advantage of interferometry is its ability to produce high 
angular resolution images, thereby allowing the  investigation of gas 
on sub-kpc to kpc scales for nearby sources.  However, spatial filtering 
by interferometers can lead to erroneous maps at large scales that could 
in turn profoundly mislead the observational interpretation of sources 
when much of the low surface brightness features are resolved out 
\citep{wilner94}. One can estimate the amount of missing flux by comparing 
the derived total flux to that of published results (if available).  

In many cases, the OVRO observations did not agree well with the flux 
measurements reported earlier by \citet{zhu99} (see Table~\ref{derived_CO}).  
Flux recovery is as low as 30\% in NGC~5395 and as high as 170\% in NGC~6621 
with an average of 82\%.   Assuming the minimum projected baseline length is 
the diameter of the dish (i.e. $\sim$ 10 m), the largest detectable structure 
is about $50''$.  This may impact some of the sources where the OVRO primary 
beam barely covers the entire  extent of  sources such as VV~219 and VV~254.   
In three cases (NGC~6621, NGC~5331S and NGC~5954) 
the interferometer flux exceeds the single dish measurements.
The discrepancy in NGC~5331S and NGC~5954 are of order 40\% and could
be attributed to the uncertainties in the (1) the amplitude calibration, 
(2) the baseline fitting for single dish measurements, and (3) the Gaussian 
(or exponential) correction factor adopted by \citet{zhu99} in order to 
estimate the flux outside the beam.  The interferometer 
flux in NGC~6621 is 70\% higher than the single dish measuments
by \citet{zhu99} where they recovered an integrated flux of
$254 \pm 29$~Jy~km~s$^{-1}$ from their fully sampled map obtained using
the IRAM~30~m telescope.
The origin of the apparent discrepancy between the single dish and
the interferometer is unclear. It may be attributed to the uncertainties 
in the amplitude calibration, 
or to spurious features in the interferometer map 
due to the uncertainties in gain calibration 
or potential erroneous data that were undetected during data editing.
Regardless of the underlying cause for the discrepancy, it is imperative 
to bear in mind that spatial filtering always exists in interferometric data, 
and interpretation of the qualitative and quantitative analysis 
should be treated with caution.   Another possible source of uncertainty 
is the use of the Galactic CO-H$_2$ conversion to estimate the 
molecular gas mass.  It has been suggested that the Galactic 
CO-H$_2$ conversion can overestimate the molecular gas mass by as 
large as factor 4 - 5 \citep{sco97,downes98} in
the nuclear starburst regions in LIRGs/ULIRGs.  
Many of our sample systems are undergoing moderate star formation activity 
(2-30~M$_\odot$ $yr^{-1}$).  However, it is 
shown in \S~6 that the CO emission is not as concentrated in the nuclear 
region as in LIRGs/ULIRGs, and therefore the use of the standard conversion is 
likely relatively safe here.

\subsection{Brightness Temperature}
The observed brightness temperature for the \ion{H}{1} emission is related
to the spin temperature by the optical depth and filling factor as
$T_B = f T_S [1-e^{-\tau}] \approx f T_S \tau$ for $\tau \ll 1$.   
It is possible that an 
edge-on galaxy could appear to have  a larger optical depth than a face-on 
galaxy due to a longer line of sight.  However, since the above measurement 
relies on the peak flux obtained in a narrow velocity channel, the 
optically thin approximation should hold true and will not introduce 
significant uncertainties in the derived temperature.  The emission 
emerging from a narrow velocity channel can be comprised of  blended 
emission from several overlapping \ion{H}{1} cloud components with 
different temperature.  The estimated spin temperature of T$_s$~$\sim 100$~K 
in a simple 3-cloud model is relatively insensitive to the 
adopted cloud densities in the optically thin limit (see Appendix).   
This implies that 
in the optically thin limit and a peak brightness temperature of 
(10 -- 20)~K (Table~\ref{derived_HI}), 
the product $\tau f$  is $\sim (0.1 - 0.2)$.  A beam filling 
factor close to unity will result in $\tau \sim (0.1 - 0.2)$, whereas a 
beam filling factor of 10\% gives $\tau \sim (1 - 2)$, implying moderate 
opacity.  It is also possible 
that the true spin temperature of these systems is higher than the above 
prediction because of a larger possible  contribution from a 
non-thermalized \ion{H}{1} cloud.  

For the CO (1--0) emission which is optically thick under most
astrophysical conditions, the peak emission in the channel maps 
can be translated into a peak 
brightness temperature through the use of the standard Rayleigh-Jeans 
approximation (Table~\ref{derived_CO}).   
The filling factor of CO~(1--0) emission is largely 
uncertain and the true brightness temperature directly scales with the 
product of filling factor and excitation temperature 
(i.e. $T_B \sim f T_{ex}$ where $f$ is the beam filling factor).  
Assuming a beam filling factor of 0.1, the measured CO~(1--0) brightness 
temperatures of 0.5 -- 2~K suggest excitation temperatures of 5 -- 20~K.

\section{Atomic and Molecular Gas Mass}
The atomic gas mass ranges from $1.8 \times 10^9 M_{\odot}$ to 
$3.4 \times 10^{10} M_{\odot}$ - a range of over one order of magnitude 
between the collision of two low mass galaxies (VV~244) and the massive 
major merger of VV~55 (Figure~\ref{MH2_MHI}~($top$)). The molecular gas 
mass is similarly diverse, ranging between $7.4 \times 10^8 M_{\odot}$ 
and $4.5 \times 10^{10} M_{\odot}$ (Figure~\ref{MH2_MHI}~($bottom$)).   
The large difference in $M_{H_2}$ can be explained by the previously known 
correlation between the size of the optical galaxy and the molecular disk 
($L_B \propto M_{H_2}^{0.72 \pm 0.03}$)\citep{young89}; i.e. in general, a 
larger stellar mass yields a larger molecular gas mass.  On the other hand, 
the \ion{H}{1} mass correlates poorly with $L_B$, possibly because a 
variety of morphological types are included in the sample.  It has been 
well known that the $M_{HI}/L_B$ ratio varies along the Hubble 
sequence \citep{roberts69}.

\begin{figure}
  \caption{Distribution of atomic (\textit{top}) and molecular (\textit{bottom}) gas mass derived from HI and CO~(1--0) observations respectively.}
  \label{MH2_MHI}
\end{figure}

The ratio $L_{IR}/M_{H_2}$ is often used to infer the star formation 
efficiency ($SFE$), with average values from 12~$L_\odot/M_\odot$ 
in isolated systems with small variation along Hubble types to 
78~$L_\odot/M_\odot$ in mergers \citep{young86} 
and some as high as a few hundred \citep{young89}.    The $SFEs$ derived 
from H$\alpha$ emission show similar results \citep{young96}.   
Using a sample of 93 galaxies, 
\citet{solomon88} found that the $SFEs$ in ``interacting pairs'' (i.e. early 
stage mergers) are comparable to those derived in isolated systems, 
suggesting the need for ``strong interactions'' (i.e. late stage mergers) 
to increase the $SFE$. They further note that 
their mean $L_{FIR}/L_{CO}$ ratios are about a factor of 2 smaller than 
those calculated from the $SFEs$ 
listed in \citet{young86}, citing differences in the methods to 
determine the single dish CO luminosities as the primary reason for the 
observed discrepancies.  The $SFEs$ in our sample are not 
particularly high ($L_{IR}/M_{H_2} = 2.7 - 15.8$~$L_\odot/M_\odot$ or
$L_{FIR}/M_{H_2} = 1.4 - 8.7$~$L_\odot/M_\odot$; 
see Table~6), and they are generally similar to those derived for 
the isolated and weakly interacting galaxies in 
\citet[][types 0, 1 \& 2 in their classification]{solomon88}.
This is a further confirmation that galaxy interaction alone cannot raise the
$SFE$, and more catastrophic events
such as a coalescence of two galaxies is needed.  There exist, however, 
systems with large $SFRs$ despite their large projected nuclear 
separation \citep[][]{trung01}.


The molecular gas mass fraction ($M_{H_2}/M_{gas}$) may determine whether 
the underlying dynamical process has a significant effect on the phase 
transition of the cold gas from atomic to molecular \citep[i.e.][]{mirabel89}.
In the present sample, 70\% of the sources 
have higher molecular gas mass than atomic gas, and this may be an 
indication that molecules dominate in colliding systems.  Past single-dish 
surveys that included a variety of galaxy morphologies have failed to reach 
a unified conclusion -- it is unclear whether tidal interaction plays a 
significant role in the phase transition \citep[e.g.][]{horellou99}.  
It is speculated that the primary discrepancy 
arises from the uncertainties pertaining to the single dish \ion{H}{1} 
and CO~(1--0) flux measurements, as well as to the validity of the Galactic  
CO -- H$_2$ conversion factor.  
Detailed analysis and discussion of the relationship 
between the \ion{H}{1} and H$_2$ gas mass content in various stages of the 
interaction will be presented as one of the main topics in Paper~III.

\section{Distribution and Kinematics of Atomic and Molecular Gas}

The CO~(1--0) and \ion{H}{1} velocity integrated line intensity 
maps overlaid on the DSS R-band images are presented in 
Figures~\ref{vv55} -- \ref{vv244} (see Table~\ref{HI_fig_prop} and 
\ref{CO_fig_prop} for the contour and gray scale level).  The maps of the 
mean velocity overlaid over the velocity dispersion are also presented 
alongside to the velocity integrated line intensity maps.  
The general features seen in these maps 
are discussed first, followed by a more detailed discussion of the 
individual sources in section \S8.  


\begin{deluxetable}{lcccccc}
\tablewidth{0pt}
\tablecaption{\ion{H}{1} Figure Properties\label{HI_fig_prop}}
\tablehead{
\colhead{Source}& \colhead{Min Contour} &  \colhead{Max Contour} & \colhead{Step}   & \colhead{Vel. Step} & \colhead{Gray Scale} \\
&  \colhead{(cm$^{-2}$)} & \colhead{(cm$^{-2}$)} & \colhead{(cm$^{-2}$)} & \colhead{(km~s$^{-1}$)} & \colhead{(km~s$^{-1}$)}   
}
\startdata
VV~55\tablenotemark{a}  & $2.0 \times 10^{20}$ & $1.0 \times 10^{21}$ & $2.0 \times 10^{20}$ & 50 & 0 - 130 \\
       & $1.0 \times 10^{21}$ & $2.6 \times 10^{21}$ & $4.0 \times 10^{20}$ &  & \\
VV~48 & $2.0 \times 10^{20}$ & $1.0 \times 10^{21}$ & $2.0 \times 10^{20}$ & 50 & 0 - 90 \\
       & $1.0 \times 10^{21}$ & $3.8 \times 10^{21}$ & $4.0 \times 10^{20}$ &  & \\
VV~254 & $2.0 \times 10^{20}$ & $2.6 \times 10^{21}$ & $4.0 \times 10^{20}$ & 50 & 0 - 200  \\
VV~253 & $2.0 \times 10^{20}$ & $1.0 \times 10^{21}$ & $2.0 \times 10^{20}$ & 50 & 0 - 120 \\
       & $1.0 \times 10^{21}$ & $2.2 \times 10^{21}$ & $4.0 \times 10^{20}$ &  &\\
VV~247 & $2.0 \times 10^{20}$ & $2.0 \times 10^{21}$ & $2.0 \times 10^{20}$ & 25 & 0 - 75\\
VV~769 & $2.0 \times 10^{20}$ & $4.0 \times 10^{21}$ & $4.0 \times 10^{21}$ & 50 & 0 - 170\\
VV~242 & $2.0 \times 10^{20}$ & $1.0 \times 10^{21}$ & $2.0 \times 10^{20}$ & 50 & 0 - 160\\
VV~219 & $2.0 \times 10^{20}$ & $1.0 \times 10^{21}$ & $2.0 \times 10^{20}$ & 50 & 0 - 37\\
       & $1.0 \times 10^{21}$ & $2.6 \times 10^{21}$ & $4.0 \times 10^{20}$ &  & \\
VV~731 & $1.0 \times 10^{19}$ & $7.0 \times 10^{19}$ & $1.0 \times 10^{19}$ & 50 & 0 - 16\\
VV~244 & $2.0 \times 10^{20}$ & $2.2 \times 10^{21}$ & $2.0 \times 10^{20}$ & 25 & 0 - 60\\
       & $1.0 \times 10^{21}$ & $6.0 \times 10^{21}$ & $8.0 \times 10^{20}$ &  &\\ 
\enddata
\tablenotetext{a}{The contour steps are increased in the inner regions 
($>10^{21}$) of some of the galaxies for better visual appearance 
of the figures.}
\end{deluxetable}

\begin{deluxetable}{lccccccc}
\tablewidth{0pt}
\tablecaption{CO Figure Properties\label{CO_fig_prop}}
\tablehead{
\colhead{Source}& \colhead{Min Contour} &  \colhead{Max Contour} & \colhead{Step}   & \colhead{Vel. Step} & \colhead{Gray Scale} & \colhead{Position Angle}\\
&  \colhead{(cm$^{-2}$)} & \colhead{(cm$^{-2}$)} & \colhead{(cm$^{-2}$)} & \colhead{(km~s$^{-1}$)} & \colhead{(km~s$^{-1}$)} & \colhead{(Degrees)}  
} 
\startdata
NGC~5257 & $2.0 \times 10^{21}$ & $2.8 \times 10^{22}$ & $4.0 \times 10^{21}$ & 50 & 0 - 130 & 270\\
NGC~5258 & $2.0 \times 10^{21}$ & $3.4 \times 10^{22}$ & $4.0 \times 10^{21}$ & 50 & 0 - 95 & 223\\
NGC~5394\tablenotemark{a} & $4.0 \times 10^{21}$ & $1.0 \times 10^{22}$ & $2.0 \times 10^{21}$ & 25 & 0 - 40 & 0\\
            & $1.0 \times 10^{22}$ & $6.0 \times 10^{22}$ & $4.0 \times 10^{21}$  \\
NGC~5395 & $2.0 \times 10^{21}$ & $1.4 \times 10^{22}$ & $4.0 \times 10^{21}$ & 50 & 0 - 40 & 0\\
UGC~12914 & $4.0 \times 10^{21}$ & $3.6 \times 10^{22}$ & $4.0 \times 10^{21}$ & 50 & 0 - 80 & 332\\
UGC~12915 & $4.0 \times 10^{21}$ & $3.0 \times 10^{22}$ & $4.0 \times 10^{21}$ & 50 & 0 - 150 & 307\\
NGC~5331 & $2.0 \times 10^{21}$ & $1.0 \times 10^{22}$ & $2.0 \times 10^{21}$ & 100 & 0 - 170 & 323\\
              & $1.0 \times 10^{22}$ & $6.0 \times 10^{22}$ & $8.0 \times 10^{21}$ \\
NGC~6621 & $2.0 \times 10^{21}$ & $4.6 \times 10^{22}$ & $2.0 \times 10^{21}$ & 50 & 0 - 114 & 309\\
UGC~813 & $4.0 \times 10^{21}$ & $3.6 \times 10^{22}$ & $4.0 \times 10^{21}$ & 50 & 0 - 45 & 315\\
UGC~816 & $4.0 \times 10^{21}$ & $1.6 \times 10^{22}$ & $2.0 \times 10^{21}$ & 50 & 0 - 50 & 0\\
NGC~7253 & $4.0 \times 10^{21}$ & $4.4 \times 10^{22}$ & $4.0 \times 10^{21}$ & 25 & 0 - 65 & 237\\
NGC~7254 & $4.0 \times 10^{21}$ & $3.0 \times 10^{22}$ & $4.0 \times 10^{21}$ & 25 & 0 - 37 & 354\\
NGC~4567 & $4.0 \times 10^{21}$ & $3.6 \times 10^{22}$ & $4.0 \times 10^{21}$ & 50 & 0 - 30 & 280\\
NGC~4568 & $1.0 \times 10^{21}$ & $6.0 \times 10^{22}$ & $4.0 \times 10^{21}$ & 50 & 0 - 45 & 204\\
NGC~7592 & $2.0 \times 10^{21}$ & $1.0 \times 10^{22}$ & $2.0 \times 10^{21}$ & 50 & 0 - 110 & 221 (257)\tablenotemark{b} \\
         & $1.0 \times 10^{22}$ & $6.0 \times 10^{22}$ & $4.6 \times 10^{21}$ & &  \\
UGC~11984 & $2.0 \times 10^{21}$ & $1.0 \times 10^{22}$ & $2.0 \times 10^{21}$ & 50 & 0 - 80 & 296\\
           & $1.0 \times 10^{22}$ & $8.2 \times 10^{22}$ & $8.0 \times 10^{21}$ & \\
\enddata
\tablenotetext{a}{The contour steps are increased in the inner regions 
($>10^{22}$) of some of the galaxies for better visual appearance 
of the figures.}
\tablenotetext{b}{() for NGC~7592E}
\end{deluxetable}

\clearpage

\begin{figure}
  \epsscale{0.7}
    \caption{Top: \ion{H}{1} map of VV~55. Middle: CO~(1--0) map of NGC~5257. Bottom: CO~(1--0) map of NGC~5258.  The corresponding mean velocity field is plotted in contours over the velocity dispersion map in gray scale on the right panel.  See Table~\ref{HI_fig_prop} and \ref{CO_fig_prop} for the contour levels}
\label{vv55}
\end{figure}

\begin{figure}
\epsscale{0.8}
    \caption{Same as Figure~\ref{vv55} but for VV~48.}
\label{vv48}
\end{figure}

\begin{figure}
\epsscale{0.8}
  \caption{Same as Figure~\ref{vv55} but for VV~254.}
\label{vv254}
\end{figure}

\begin{figure}
    \caption{Same as Figure~\ref{vv55} but for VV~253.}
\label{vv253}
\end{figure}

\begin{figure}
   \caption{Same as Figure~\ref{vv55} but for VV~247.}
\label{vv247}
\end{figure}

\begin{figure}
  \caption{Same as Figure~\ref{vv55} but for VV~769.}
\label{vv769}
\end{figure}

\begin{figure}
    \caption{Same as Figure~\ref{vv55} but for VV~242.}
\label{vv242}
\end{figure}

\begin{figure}
\epsscale{0.8}
    \caption{Same as Figure~\ref{vv55} but for VV~219.}
\label{vv219}
\end{figure}

\begin{figure}
    \caption{Same as Figure~\ref{vv55} but for VV~731.}
\label{vv731}
\end{figure}

\begin{figure}
\epsscale{0.8}
    \caption{Same as Figure~\ref{vv55} but for VV~244.}
\label{vv244}
\end{figure}

\subsection{\HI\ and CO Morphology} 

The summary of \HI\ and CO~(1--0) morphological properties 
in Table~\ref{morph_prop} 
highlights some of the distinct features seen in the maps presented in
Figures~\ref{vv55} - \ref{vv244}.
Specifically, we identify systems with visual signature of
(1) a long \HI\ tail, (2) substantial \HI\ emission in the medium between
the galaxies in the pair, (3) an offset between \HI\ peaks 
and the stellar light, (4) an offset between CO~(1--0) peaks and the
galaxy center traced in K-band emission, and (5) an isolated CO~(1--0) 
emission with no apparent optical counterpart.
A short discussion of each feature and the galaxies identified are 
presented in what follows.

N-body simulations of galaxy interactions suggest that ejection of 
cold disk gas at large radii, primarily neutral hydrogen,  
into tidal tails is a ubiquitous phenomenon.  The degree of 
ejection depends crucially on the encounter geometry 
\citep[see ][]{toomre72}.  
Tidal ejection is expected to occur to some degree even in an orbital 
geometry that is the least favorable to transfer energy and 
angular momentum (i.e. a retrograde encounter). 
Long and well defined tidal tails are unlikely to form in such an event.  
Inspecting the size and the amplitude of the tidal tails, therefore, 
offers a unique way to infer the orbital geometry 
\citep{hibbard96}.  Out of the nine systems with high quality \ion{H}{1} 
maps, five (VV~55, VV~253, VV~247, VV~731, VV~244) 
display long \ion{H}{1} tidal tails (i.e. $D_{HI}/D_{25} > 1$) 
suggesting that at least one of the disks involved is in a prograde 
orbit (Table~\ref{morph_prop}).  The highly inclined geometry complicates 
the analysis in VV~242 despite hints of \ion{H}{1} tails seen at the edges 
of both disks.  It is not surprising to find that about 50\% of the 
sources are in a prograde orbit 
since there are two equally possible disk orientations, 
unless the encounter geometry is such that it is a head-on collision 
similar to  the ``Cartwheel galaxy (VV~784)''. 

The occurrence of a long \HI\ tail appears to coincide with the elevated
level of infrared emission -- out of the five systems with long \HI\ tails,
four (80\%) are classified as LIRGs.  Correlation between other morphological
properties with infrared activity appears to be relatively 
insignificant (see Table~\ref{morph_prop}).  A strong perturbation from
a prograde orbit can lead to efficient ejection of the outer disk
gas into tidal tails, while the inward propagating waves can compress and heat
the inner disk gas, simultaneously inducing intense star 
formation activity there.
While the presence of a long \HI\ tail may infer 
the interaction history and the increased bursts of star 
formation in a collision of gas-rich 
progenitor galaxies, it is possible that the strength of the
interaction and the associated 
infrared activity are correlated with other important physical parameters. 
These include, for example, the temperature of the dust \citep{xilouris04} 
and/or the presence of the nuclear AGN.
It is, however, suggested that AGN dominated galaxies constitute only of
order (1--2)\% of the infrared luminous galaxy population \citep{yun01b}.
Further discussion 
on this apparent correlation will be conducted in Paper~III.


A recent, strong head-on collision can lead to a substantial amount
of radio continuum emission in the medium between the two galaxies
as seen in the two Taffy systems (see \S7, \S8.3, \S8.6).  A long
stretched morphology is also seen in the \HI\ emission in these 
galaxies.  Therefore, substantial \HI\ emission in the medium between
the two galaxies may signify a strong
and recent encounter.  By searching for systems in which 
the $N_{\rm HI} = 10^{21}$~cm$^{-2}$ contours are connected between the
two galaxies, nearly 80\% (7/9) of the systems show visual evidence 
of such feature.  This may be a slight overestimate
as the line of sight projection can significantly impact the apparent 
\HI\ morphology in some cases (i.e. VV~242, VV~244).  
In Paper~III, this feature will be used as one of the important 
criteria to define the merger chronology of the sample.

A transient and inelastic response of the gas is particularly noticeable
during the initial stages of a disk-disk collision in numerical
simulations, sometimes resulting in a strong asymmetry in the gas distribution 
(see Paper~I).  In addition to the displacement of the CO peaks
from the stellar light (see below), a significant displacement of \HI\ peaks
from the stellar disks is also seen among six (VV~244, VV~247, 
VV~253, VV~254,  VV~731 and VV~769) of the systems.  Such displacements
are generally not expected since gas and stellar structures inside
the tidal radius should remain unaffected by a tidal disruption.  
A pre-existing asymmetry or a central hole in the gas distribution 
prior the recent collision may account for the observed displacement
in some cases.  An inelastic gaseous collision is primarily responsible
for the gas morphology 
in the two ``Taffy'' systems (VV~254 and VV~769), and it may play an
important role in shaping the gas distribution in many of the other
systems as well.  

In general, a smooth and continuous distribution of CO~(1--0) in the 
disk is seen, but 
clumpy distributions are evident in a few sources that are gas poor 
(UGC~12914, UGC~816, NGC~4567, NGC~5395).  CO~(1--0) emission is 
fully resolved spatially with at least 3 synthesized beams across the 
stellar disks in all cases, widely extended with respect 
to the R-band emission at the column density of 
$2.0 \times 10^{21}$ cm$^{-2}$.  This is in a stark contrast to
the observations of IR luminous, more advanced interaction/merger
systems whose CO emission is typically characterized by a compact 
($\le1$ kpc) nuclear concentration centered on the local minima
in gravitational potential \citep[][]{sco97,downes98,bryant99,yun01}.

The peak of the CO~(1--0) emission does not coincide with the 
dynamical center of the galaxy determined from K-band emission 
in NGC~5395, NGC~5258, UGC~12914, or NGC~5331N.  There are 
five CO  emitting complexes with no optical counterpart (UGC~12914, 
UGC~816, NGC~5331S, NGC~5954 and NGC~6621).  The true nature of these 
complexes is yet to be determined, but one possible explanation is the 
ejection of molecular gas from strong gas-gas collisions 
during the first pericentric passage.

Finally, the galaxy environment seems to play a relatively 
minor role in the formation 
and the shaping of the long \ion{H}{1} tails.  
While previous \HI\ observations with much lower surface brightness 
sensitivity have shown that galaxies in a 
dense cluster environment harbor less extended atomic gas than galaxies in 
isolation \citep{cayatte90}, such pattern is not obvious in the current data 
(i.e. long tidal tails are seen independent of the environment, with a
possible exception of VV~219 in Virgo).  Similarly, 
whether the environment affects the distribution and the physical 
properties of the CO~(1--0) emitting clouds is uncertain from our data.

\begin{deluxetable}{lccccccc}
\tabletypesize{\scriptsize}
\tablewidth{0pt}
\tablecolumns{10}
\tablecaption{\ion{H}{1} and CO(1--0) Morphological Properties\label{morph_prop}}
\tablehead{
\colhead{Source}& \colhead{$L_{\rm FIR}$} & \colhead{No. of} & \colhead{Long HI tail\tablenotemark{b}} & \colhead{Stretching HI\tablenotemark{c}}  & \colhead{HI peaks\tablenotemark{d}} & \colhead{CO~(1--0) peak\tablenotemark{e}} &\colhead{Isolated CO~(1--0)\tablenotemark{f}}\\
 & (10$^{10}$ L$_{\odot}$) &\colhead{Neighbors\tablenotemark{a}} & & &\colhead{$\neq$ stellar light} & \colhead{$\neq$ galaxy center} &
} 
\startdata
VV~55      &28.3&12 & Y~(1.1) & Y &N \\
~~~NGC~5257&& & & & &N & N \\
~~~NGC~5258&& & & && Y~(6) & N \\
VV~48      &6.1&23 & N~(0.9) & N &N\\
~~~NGC~5394&& & & && N & N \\
~~~NGC~5395&& & & && Y~(7) & N\\
VV~254     &7.0&6 & N~(0.9) & Y &Y\\
~~~UGC~12914&&& & && Y~(8) & Y~(9.2)\\
~~~UGC~12915&&& & && N & N  \\
VV~253     &32.3&17& Y~(2.0) & Y &Y\\
~~~NGC~5331N&&& & & &Y~(3) & N  \\
~~~NGC~5331S&&& & & &N & Y~(9.0) \\
VV~247      &14.2&1 & Y~(1.2) & Y &Y\\
~~~NGC~6621 &&& & && N & Y~(9.5)  \\
~~~NGC~6622 &&& & & &\nodata & \nodata  \\
VV~769      &4.5&1& N~(0.9) & Y &Y\\
~~~UGC~813  &&& & && N & N  \\
~~~UGC~816  &&& & & &N & Y~(9.0)  \\
VV~242      &7.1&1& N~(0.7) & Y  &N\\
~~~UGC~11984&&& & && N & N  \\
~~~UGC~11985&&& & & &\nodata & \nodata  \\
VV~219      &2.2&58& N~(0.5) & N  &N\\
~~~NGC~4567 &&& & && N & N  \\
~~~NGC~4568 &&& & && N & N  \\
VV~731      &21.5&0& Y (\nodata) & \nodata  &Y\\
~~~NGC~7592W&&& & && N & N \\
~~~NGC~7592E&&& & & &N & N  \\
VV~244      &2.2&20& Y~(1.0) & Y&Y\\
~~~NGC~5953 &&& & && N & N  \\
~~~NGC~5954 &&& & && N & Y~(8.0)\\
\enddata
\tablenotetext{a}{The number of neighboring sources found in Table~\ref{environment}.}
\tablenotetext{b}{A long HI tail is defined when $D_{HI}/D_{25}$ exceeds unity (numbers shown in ()). $D_{HI}/D_{25}$ is the ratio between the major axis of HI and the sum of the 25th isophote of the B-band image from RC3.  The HI major axis is estimated from from the lowest column density ($N = 2 \times 10^{20}$ cm$^{-2}$) contours.  $D_{HI}/D_{25}$ is not calculated for VV~731 because of the poor quality of the HI data.}
\tablenotetext{c}{The stretching HI emission is defined when the $10^{21}$ cm$^{-2}$ column density contours of the pairs appear visually connected.}
\tablenotetext{d}{Identified systems with significant displacement of \HI\ peaks from the stellar disk.}
\tablenotetext{e}{When the location of the CO~(1--0) peak column density is not consistent with the center of the galaxy determined from the K-band image to within the CO~(1--0) angular resolution.  The offsets are shown in () in kpc.}
\tablenotetext{f}{Identified  CO~(1--0) complexes with no obvious visual evidence of associated optical emission.  The masses of the CO~(1--0) complexes are shown in () in logarithmic scale of $M_{\odot}$.}
\end{deluxetable}

\subsection{Position Velocity Diagrams} 

The Position Velocity Diagram (PVD) is a commonly used tool for 
inferring the gas kinematics and the rotation of galaxies.  The 
KPVSLICE routine in KARMA \citep{gooch95} allows interactive construction 
of the PVDs using a 3 dimensional datacube as an input.  The coordinates of 
the K-band emission peak was used to align the center of the PVD slit 
(see Table~\ref{galaxy_sample}), which was then rotated interactively until
 the slit position angle matched the apparent kinematic major axis.   For 
consistency, the position angle of the slit was chosen to lie in the eastern 
half for every galaxy, and thus the orientation of the resulting PVDs are 
different depending on the direction of the disk rotation along the line 
of sight.  

The results are shown in Figure~\ref{pvd1} -  \ref{pvd3}, and the adopted 
position angles listed in Table~\ref{CO_fig_prop}.  The morphology of 
the CO~(1--0) PVDs varies significantly depending on the amount of 
disk gas, spatial resolution, and the presence of tidally induced 
non-circular motion dominating both the inner and the outer parts of the 
interacting galaxies.  The dominance of symmetric morphology in the
PVDs among nearby isolated galaxies \citep[see][]{sakamoto99} 
contrasts strongly with the 
CO~(1--0) PVDs shown in Figure~\ref{pvd2} -  \ref{pvd3} where
about one half are asymmetric with respect to the dynamical center. 
A close examination of the \ion{H}{1} and CO~(1--0) PVDs reveals that 
the peak of the emission is radially offset toward one or 
both directions from the PVD centroid (K-band peak) in many cases.   
Absorption can affect the observed \ion{H}{1} distribution  
to some degree, but opacity effect is less important for CO~(1--0).   
Among the PVDs that display smooth and continuous CO~(1--0) distribution 
with high enough S/N, more than 50\% (NGC~4567/8, NGC~5953/4, NGC~7592, and 
UGC~12914/5) show this anomalous kinematic signature in CO~(1--0) emission. 
The simulation analysis presented in Paper~I suggests that 
a central depression in the molecular gas emission and the displacement
of the emission peak  
may indicate the existence of a 
central gas ring or tightly wound spiral arms.  The existence 
of periodic orbits (or a stellar bar) is one possible explanation for the 
formation of a central ring.   Such a structure does not directly imply gas 
inflow, but it suggests that gas orbits possess axial 
symmetry such as a ring or ``twin peaks'' \citep{kenney92} on kpc scales.  
Using images with angular resolution that is a factor of a few better 
\citep{sakamoto99}, these peculiar structures were found  to dominate the 
kinematics in some of the isolated galaxies on scales of few hundred 
parsecs. Angular resolution  limitations may affect the interpretation of 
these structures to some degree for our program sources. 

\begin{figure}
    \caption{HI and CO~(1--0) PVDs.  The center of the PVD is chosen to 
coincide with the 2MASS K-band peak in each galaxy, and 
the slit is interactively aligned to match the stellar disk major axis.  
The position angles of the slit are listed in Table~\ref{CO_fig_prop}.
The range in the angular offset (x-axis) is significantly different 
between \HI\ and CO~(1--0) because they are different gas tracers.  The
velocity ranges (y-axis) are similarly different in some cases when the
\HI\ and CO~(1--0) kinematics are inconsistent. 
}
\label{pvd1}
\end{figure}

\begin{figure}
    \caption{continue from Figure~\ref{pvd1}.  The CO~(1--0) PVD for NGC~6622
and UGC~11985 are not available (see text).}
\label{pvd2}
\end{figure}

\begin{figure}
    \caption{continue from Figure~\ref{pvd2}. The \HI\ PVD for NGC~7592 is 
not available (see text).}
\label{pvd3}
\end{figure}

The simulation analysis in Paper~I demonstrates that an emission in the 
forbidden velocity quadrants of the PVD indicates a radial motion such 
as gas inflow.  There are two cases in this sample in which clear
signatures of substantial CO~(1--0) emission in the 
forbidden velocity quadrants are seen: UGC~12915 and NGC~6621. 
This peculiar emission in UGC~12915 is seen near the center of the 
galaxy ($\sim 1 - 2$ kpc from the K-band peak), and the associated 
molecular gas mass is $9 \times 10^8 M_\odot$, which is about 10\% of the 
total molecular gas mass of UGC~12915.  A similar feature is seen in 
NGC~6621, and it is associated with the southeastern CO~(1--0) complex 
about 6~kpc south of the NGC~6621 nucleus.  
The emission from this complex may belong to the 
southern tidal arm and to the star forming clusters developing near the 
overlapping region.  If  the southern CO~(1--0) complex in NGC~6621 is 
involved in a radial inflow along the tidal arm, this indicates that about 
10\% of the sources detected in CO~(1--0) is showing a possible evidence 
of inflow, consistent with the predictions made in Paper~I.

\subsection{Rotation Curves}

The rotation curve fitting was performed by tracing the emission envelope
of the \ion{H}{1} and CO~(1--0) PVDs corrected for the instrumental velocity 
resolution (see Paper~I).  While the turbulence term removes 
$\sim 10$ km s$^{-1}$ from the rotation velocity in normal spiral galaxies, 
this term is neglected here for consistency since the contribution 
from turbulent motion in the ISM of colliding systems is unknown.  Both 
sides of the fitted rotation curves are averaged to derive the 
final rotation curve.   Furthermore, to alleviate the discrete velocity 
sampling and the clumpy nature of the gas emission, each derived rotation 
velocity is smoothed with a Gaussian whose width is equal to 
the FWHM of the synthesized beam.  Lastly, to avoid oversampling the PVD 
fit, the number of points were adjusted, typically by a few points per beam,  
in order to present the points in a smooth and continuous manner.  The 
resultant rotation curves are presented in Figures~\ref{rot1} -- \ref{rot3}. 

\clearpage

\begin{figure}
    \caption{The fitted rotation curves using HI (\textit{open circles}) and CO~(1--0) (\textit{filled circles}) data for VV~55, VV~48, VV~254 and VV~253.  Units are arc minutes (\textit{bottom}) and kpc (\textit{top}) for the x-axis and km s$^{-1}$ for the y-axis.   The dashed (CO~(1--0)) and solid (HI) error bars represent the velocity (y-axis) and spatial (x-axis) resolution.}
\label{rot1}
\end{figure}

\begin{figure}
    \caption{Similar to Figure~\ref{rot1} but for VV~247, VV~769, VV~242 and VV~219.}
\label{rot2}
\end{figure}

\begin{figure}
    \caption{Similar to Figure~\ref{rot1} but for VV~731 and VV~244.}
\label{rot3}
\end{figure}

The high resolution OVRO CO~(1--0) data are used to describe the inner 
rotation whereas the outer parts of the disk rotation is determined from 
the lower resolution \ion{H}{1} data.  Determination of the nuclear gas 
rotation is limited by the 
angular resolution of the CO~(1--0) observation ($\sim 1$ kpc), and thus 
the kinematics of the rising part of the rotation curve cannot be derived 
with a high accuracy.  The rotation curves in the outer parts of the disks 
are generally flat (within 10 - 20\%) with only a few cases showing 
significant warps (UGC~813/6 and NGC~5257).  A smooth transition from the 
CO~(1--0) rotation to \ion{H}{1} rotation is seen in 10 of the 15 sources 
that had robust detection in both \ion{H}{1} and  CO~(1--0).  Sources that 
fail to exhibit smooth and continuous transitions are primarily those with 
low abundance of the CO~(1--0) emitting clouds (e.g. UGC~813/6) although
spatial segregation between \HI\ and CO may be important in some cases
(e.g. NGC~5394, NGC~5953/4).  A high S/N detection of both 
CO~(1--0) and \ion{H}{1} results in a well determined rotation curve, 
and this proves the robustness of our fitting algorithm.  

The derived rotation curves can now be used to derive 
the dynamical mass of each 
galaxy.  The dynamical mass is determined from $M_{dyn} = \frac{R ~V^2}{G}$ 
where $R$ is the radius of the detected CO~(1--0) emission, 
$V = V_{rot}/\sin i$, where $i$ is the disk inclination, and $G$ is the 
gravitational constant.  Because the disk inclination in interacting systems 
is exceedingly difficult to determine, we assume $i = 90$ yielding a lower 
limit in all cases.  In addition, $M_{dyn}$ is only computed out to the 
maximum radial extent of the CO~(1--0) emission since the kinematics of 
the outer disk may be dominated by the tidally perturbed \ion{H}{1} gas and  
may not represent the true rotation of the galaxy.  The coarse angular 
resolution of the \ion{H}{1} emission further complicates the exact 
identification of the diffuse component of the ISM.   The results presented in 
Table~\ref{derived_CO} show a wide variety of $M_{H_2}/M_{dyn}$ ranging from 
0.02 in NGC~5395 to 1.58 in NGC~5394.  NGC~5394 is nearly face on and hence 
the dynamical mass is underestimated substantially (by $\sin^2 i$).  
For the two edge-on galaxies UGC~12915 and UGC~11984, an inclination 
correction is not necessary.  The $M_{H_2}/M_{dyn}$ 
ratios of these two systems are 0.18 and 0.14 respectively, and they are 
consistent with the observed ratios in the inner 500~pc of  barred spiral 
galaxies \citep{sakamoto99}.  It is important to bear in mind that 
non-circular motion from the bar can result in an over-estimate of 
the dynamical mass, but the analysis of numerical simulations suggests that 
the uncertainty is limited to 30 to 50\% (Paper~I).

\section{Radio Continuum}

The 1.4~GHz radio continuum is obtained by combining the line free channels 
in the \ion{H}{1} spectroscopic data.  The maps and the total flux are 
presented in Figure~\ref{radio_cont} and Table~\ref{radio_cont_tab} 
respectively.  Natural weighting was used in all images in order to maximize 
the surface brightness sensitivity.  The two galaxies in seven of the 
systems are clearly resolved, and the total fluxes were derived separately 
in Table~\ref{radio_cont_tab}.  For those that were not completely resolved, 
the total flux of the two galaxies is given. 

Substantial amount of radio continuum in the region between the two galaxies 
signifies a recent and head-on encounter \citep{cond93, cond02}.  Out of the 
ten systems, only the previously known Taffy systems (VV~254, VV~769) show 
an obvious radio continuum bridge that connects the two galaxies.  
The radio continuum peaks in the overlap region between the two galaxies
in VV~253, but the poor angular resolution of the data makes it
difficult to draw a firm conclusion.  Some of the other systems may 
have also undergone a head-on collision similar to those seen 
in the Taffy systems, but projection will likely impact the apparent 
morphology of the radio continuum images.  Without the source of
energy for re-acceleration of electrons, the radio bridge would
quickly fade away, and only the systems that have undergone a 
collision within the last few tens of million years can be 
identified in radio continuum.

The SFRs are derived from the radio continuum fluxes for the whole system, 
and separately for the systems when resolved (Table~\ref{radio_cont_tab}).  
There are three systems (VV~55, VV~253, VV~731) in which the 
$\rm SFR_{1.4GHz}$ is noticeably larger than the rest.  Even though the radio 
derived SFR, which indirectly translates to the level of 
\textit{current} star formation 
activity, is relatively low compared to those typical of ULIRGs 
(SFR = 100-1000~M$_\odot$ yr$^{-1}$), 
the large amount of molecular gas found in 
two of these systems (VV~55, VV~253)($\sim 10^{10}$ M$_{\odot}$) may be 
enough to initiate and sustain future bursts of star formation.  Molecular 
gas mass is low ($< 10^{10}$ M$_{\odot}$) in VV~731 despite its relatively 
high SFR.

\begin{figure}
    \caption{The distribution of the 1.4 GHz radio continuum emission from all of our program sources.  The contours are 1,10,30,50,70,90\% of the peak flux.}
\label{radio_cont}
\end{figure}

\begin{deluxetable}{lrr}
\tabletypesize{\small}
\tablewidth{0pt}
\tablecaption{The 1.4 GHz Flux and Star Formation Rates\label{radio_cont_tab}}
\tablehead{
\colhead{Source}&\colhead{$S_{\rm 1.4GHz}$} &  \colhead{SFR$_{\rm 1.4GHz}$\tablenotemark{a}} \\ 
 & ($Jy$) & ($M_\odot$ $yr^{-1}$) 
} 
\startdata
VV~55  &  & \\
~~~NGC~5257 & 0.048 & 27.8 \\
~~~NGC~5258 & 0.043 & 24.9 \\
VV~48  &  & \\
~~~NGC~5394 & 0.033 & 4.2 \\
~~~NGC~5395 & 0.059 & 7.4 \\
VV~254 & &\\
~~~UGC~12914 & 0.018 & 3.9\\
~~~UGC~12915 & 0.043 & 9.3\\
VV~253 & 0.044 & 53.9 \\
VV~247 & 0.030 & 14.1 \\
VV~769 & & \\
~~~UGC~813 & 0.018 & 5.8\\
~~~UGC~816 & 0.022 & 7.1\\
VV~242 & & \\
~~~NGC~7253 & 0.070 &16.2\\
~~~NGC~7254 & 0.009 &2.1\\
VV~219 & &\\
~~~NGC~4567 & 0.010 &0.6\\
~~~NGC~4568 & 0.122 &6.8\\
VV~731 & 0.075 & 51.4 \\
VV~244  & &\\
~~~NGC~5953 & 0.074 & 3.0 \\
~~~NGC~5954 & 0.021 & 0.9 \\ 
\enddata
\tablenotetext{a}{Using SFR ($M_\odot$~yr$^{-1}$) = 
$5.9 \times 10^{-22}$~L$_{\rm 1.4 GHz}$~(W~Hz$^{-1}$) \citep{yun01b}}
\end{deluxetable}

\section{Individual Sources}

In this section, some background information on the individual sources is provided, and a short description of the features seen in the atomic and molecular gas maps and the PVDs are offered.

\subsection{VV~55 (NGC~5257/8, UGC~8641, ARP~240)}

At a distance of 90 Mpc, VV~55 is one of the more distant sources in the 
sample of objects presented here.  It is classified as a LIRG from its high 
infrared luminosity ($L_{IR} = 2.8 \times 10^{11}$~L$_{\odot}$).  NGC~5257 
is a late type spiral with an inclined ``S'' optical morphology, where the 
northern part of the ``S'' shows a western tidal extension 18 kpc 
long.  The southern tidal feature points eastward and is connected to the 
northern arm of its companion galaxy, NGC~5258. A low surface brightness 
diffuse arm 13~kpc long emerges from the southern part of the ``S'' 
pointing directly south.  Detailed features in the nuclear region of  
NGC~5257 cannot be seen due to extinction and poor angular resolution of 
the DSS image, but there is a hint of a dust-lane along the tidal arms. 
NGC~5258 also shows two tidal features forming a slightly more vertically 
elongated ``S''.  The southern arm is 40~kpc in length, much longer 
than the extent of the northern arm.  The R-band emission 
in the nuclear region shows patchy features suggesting obscuration by dust.  
The H$\alpha$ emission 
map shows high levels of star formation activity in the southern arm of 
NGC~5257, and some activity in the region southwest of the nucleus in 
NGC~5258, while the nuclei of both galaxies show a low degree of star 
formation activity \citep{bush90}.   The far infrared emission at 100 and 
160 $\micron$ obtained with the Kuiper Airborne Observatory (KAO) shows 
dust emission emerging from both galaxies \citep{bush98}.

The \HI\ distribution and kinematics, imaged using the VLA at
a $22.2''\times 17.6''$ ($9.8 \times 7.8$ kpc) resolution, are shown 
in Figure~\ref{vv55}.
The relatively undisturbed distribution of the \ion{H}{1} emission 
in NGC~5257 (Figure~\ref{vv55} (\textit{top})) suggests that either most 
of the large scale tidal disturbance occurs along the line of sight, or 
this particular orbital geometry is less susceptible to the formation of 
large tidal features (i.e. a retrograde encounter).  The diffuse \ion{H}{1} 
emission south of the galaxy shows a small lump that might be associated 
with the low surface brightness optical tail, in which case the structure 
may grow to a characteristic \ion{H}{1} tail in the future.  In contrast, 
the more face-on galaxy NGC~5258 shows an \ion{H}{1} tail 60~kpc long, 
extending beyond the optical tail traced in the DSS image. This is probably 
an indication of a prograde encounter.  Two peaks separated by 11~kpc can 
be identified at the northern and southern edge of the disk.  The southern 
peak is spatially coincident with the star forming region found in the 2MASS 
and H$\alpha$ images.  

The CO~(1--0) distribution and kinematics, imaged using the OVRO array at
a $6.2''\times 4.3''$ ($2.7 \times 1.6$ kpc) resolution, are shown 
in Figure~\ref{vv55}.
The CO distribution in NGC~5257 (Figure~\ref{vv55} (\textit{middle})) 
shows a concentration near the nucleus with emission extending both in the 
northeast and southwest forming a 10~kpc long bar.  The southern 
arm shows slightly brighter CO~(1--0) emission and is possibly related to 
the star forming region detected in H$\alpha$ emission there.  
The CO~(1--0) emission in 
NGC~5258 (Figure~\ref{vv55} (\textit{bottom})) forms an elongated ``S'' 
shape that closely traces the similar optical morphology.  
High concentration of CO~(1--0) emitting clouds is seen toward the 
southern portion of the ``S'', consistent with the elevated level 
of H$\alpha$ emission in the same region \citep{bush90}.
This is one of 
the few examples in this sample of 10 interacting systems where the 
CO~(1--0) emission traces  the tidal arms seen in the optical image.   
This is a possible indication of an early stage interaction as such a
gas morphology is seen during the inflow 
phase ($t \leq 10^8$ years) in the numerical simulations (Paper~I).   

The peak intensity in the \ion{H}{1} PVD (Figure~\ref{pvd1}) of NGC~5257 is 
offset from the center of the galaxy by $4 - 8$~kpc while the CO~(1--0) 
emission in the PVD (Figure~\ref{pvd1}) is more uniform.  The 
lopsided \ion{H}{1} distribution is also evident in the velocity 
integrated line intensity map (Figure~\ref{vv55} (\textit{top})),
and this is possibly due to higher abundance of CO~(1--0) emitting clouds
in the warmer and denser conditions near the central region of the galaxy.  
The derived rotation curve (Figure~\ref{rot1}) 
connects smoothly from CO~(1--0) to \ion{H}{1} measurements at 
$V_r \sim 250$~km~s$^{-1}$.  
The rotation curve declines quickly in velocity 
($\Delta v > 100$~km~s$^{-1}$) at large radii ($>15$~kpc), 
indicating that the outer \ion{H}{1} disk is strongly warped
by tidal stripping of the loosely bound outer disk gas.  
In NGC~5258, the peak of the \ion{H}{1} PVD is 
lopsided toward the negative offset by 9~kpc while the CO~(1--0) PVD is 
also lopsided but in the opposite direction.  The rotation curves 
derived from these PVDs agree to within 50~km~s$^{-1}$,
with a slightly lower velocity for CO. 
In contrast to the declining velocity seen at large radii in 
NGC~5257, the \ion{H}{1} rotation in NGC~5258 maintains a consistent 
velocity at 200 - 250~km~s$^{-1}$ out to 50~kpc.

\subsection{VV~48 (NGC~5394/5, UGC~8898, ARP~84)}

VV~48 consists of two late type spiral galaxies, NGC~5394 and NGC~5395,
with a projected nuclear separation of 26~kpc.  
NGC~5394 harbors a strong central starburst \citep{sharp85, kaufman99} 
with 10~kpc long tidal arms in both northern and southern directions.  
NGC~5395 also has multiple tidal arms, with enormous dust lanes dominating 
the western arm of the galaxy.  Extensive star formation activity is seen 
in H$\alpha$ emission along the spiral arms forming a large scale elliptical 
star forming ring \citep{kaufman99}.   A detailed study of the \ion{H}{1} 
observations using the VLA, a numerical model of the encounter, 
and BIMA CO~(1--0) observations of the northern part of the system are 
presented in \citet{kaufman99} and \citet{kaufman02}.  
The BIMA observations had two pointing centers; one toward NGC~5394 
and the other toward the northern half of NGC~5395.  
Kaufman et al. reported that 
80\% of the CO~(1--0) emission in NGC~5394 is found near the central 
starburst and that NGC~5394 orbit is prograde with respect to the orbital 
geometry.  The ``eye-shape'' morphology of NGC~5395 prompted them to call 
this system in a ``post-ocular phase''.

The \HI\ distribution and kinematics, imaged using the VLA at
a $17.8''\times 16.3''$ ($4.0 \times 3.6$ kpc) resolution and shown 
in Figure~\ref{vv48}, are the results of our own reduction of the
archival data already presented by \citet{kaufman99}.
Multiple \ion{H}{1} peaks trace the western dust lane in NGC~5395 
while a lower column density gas fills the central and eastern 
portion of the disk (see Figure~\ref{vv48} (\textit{top})).  
Yet fainter \ion{H}{1} extension to the north reaches 15~kpc beyond 
the edge of the stellar disk, and it has a faint optical counterpart
in a deep R-band image \citep{kaufman99}. 
One possible explanation to the deficiency of \HI\ in NGC~5394 
\citep[$M_{HI} = 7.3 \times 10^8$;][]{kaufman99} is the disk-wide phase 
transition of atomic to molecular gas.  This is evidenced from the large amount
of disk molecular gas mass ($M_{\rm H_2} = 3.2 \times 10^9$~M$_{\odot}$), 
as well as the strong H$\alpha$ and 
radio continuum emission signifying disk-wide star formation 
activity in NGC~5394 \citep{kaufman99}.

The CO~(1--0) distribution and kinematics, imaged using the OVRO array at
a $5.5''\times 4.3''$ ($1.2 \times 1.0$ kpc) resolution, are shown 
in Figure~\ref{vv48}.  The new CO data presented here are not only 
more complete in its coverage but they also represent a three fold improvement
in sensitivity and a 20\% improvement in resolution over the earlier
BIMA observation by \citet{kaufman02}.  Strong CO~(1--0) 
emission is seen in NGC~5394 with a slightly lopsided distribution toward 
the southwestern side of the galaxy (Figure~\ref{vv48} (\textit{middle})). 
The shallow velocity gradient across the CO~(1--0) emitting region 
suggests that the galaxy is nearly face-on.  The low column 
density CO~(1--0) extension to the east was 
not seen in the lower sensitivity BIMA map \citep{kaufman02}.  
The clumpy distribution of CO~(1--0) emission 
in NGC~5395 (see Figure~\ref{vv48} (\textit{bottom})) roughly traces the 
star forming ring seen in H$\alpha$ emission. Each of the four large CO
clumps has mass of  $M_{H_2} = (0.5 - 1.0) \times 10^9 M_{\odot}$.  
The northern CO clump was only marginally detected in the less 
sensitive BIMA observation, whereas the southern portion of the galaxy 
was not covered by \citet{kaufman02}.
This is the only galaxy in which CO~(1--0) emission is completely 
absent in the nuclear region among all sample sources detected in CO 
-- a 3$\sigma$ upper mass limit for the nuclear molecular gas complex
is $5\times 10^6 M_\odot$.  
The two southwestern CO clumps coincide with the peaks in 
\ion{H}{1}, but the two CO clumps in the eastern half of NGC~5395 
have no \ion{H}{1} counterparts.  The central CO clump, 
in particular, is detected where the \ion{H}{1} emission has a local 
minimum ($N_{HI}\sim 10^{21}$ cm$^{-2}$).  Because 
of its clumpy nature, tracing the rotational kinematics in NGC~5395 using 
the CO~(1--0) emission is difficult (Figure~4 (\textit{bottom right})). 
Nevertheless, the north-south velocity gradient characteristic of
the disk rotation is quite evident.  The comparison of the \HI\
and CO mean velocity fields (Figure~\ref{vv48} 
(\textit{top right}) with Figure~\ref{vv48} (\textit{bottom right}))
suggest that the velocities of 
the CO emitting clouds are roughly consistent with the velocities of 
the \ion{H}{1} clouds at the same location in the galaxy. 

The \HI\ emission in NGC~5394 is very faint, but the \ion{H}{1} 
PVD (Figure~\ref{pvd1}) clearly shows a disk-like material
orbiting the galaxy.  A careful comparison with the CO
PVD (Figure~\ref{pvd1}) shows that the velocity 
gradient is much shallower for the \HI, suggesting that the 
neutral atomic gas lies much farther outside the CO disk,
perhaps forming a partial ring or a tidal tail.
The CO~(1--0) kinematics in NGC~5394 cannot be investigated in detail 
because of the limited angular resolution.  
The face-on orientation of the galaxy results in a very low rotational 
velocity ($V_r\sim 50$ km~s$^{-1}$) in both CO~(1--0) and \ion{H}{1} 
(see Figure~\ref{rot1}).  The \ion{H}{1} 
PVD in NGC~5395 shows peaks near both sides of the tip of the rotation, 
with a depression of \ion{H}{1} toward the galaxy center.  
The double peak morphology arises from the higher column density 
\ion{H}{1} that is organized in a large scale ring with a 
diameter $\sim 24$~kpc.  
The low abundance of CO emitting clouds in NGC~5395 makes the 
PVD appear clumpy and discontinuous, but the correspondence to the
\HI\ PVD is quite good.  The derived CO rotation curve is slightly
lower than that of \HI\ (Figure~\ref{rot1}), possibly due to the 
complexity of the fit arising from the clumpy and faint nature of the
CO emission.

\subsection{VV~254 (UGC~12914/5, Taffy~I)}
The interacting galaxy pair, VV~254 (hereafter Taffy I), was studied in 
\ion{H}{1} and radio continuum by \citet{cond93}.  They found \ion{H}{1} 
and radio continuum emission connecting the two galaxies whose morphology 
resembles that of stretching bands of a taffy candy.  The long stretched 
morphology seen in both \ion{H}{1} and radio continuum suggests that the 
collision was strong enough to pull substantial amounts of gas out of the 
disk, and that the collision was recent enough for the high energy electrons 
to be still radiating in synchrotron emission.  Spectral steepening in 
the taffy medium constrains the merger age to within  
$(1-2) \times 10^7$ years of the initial collision \citep{cond93}.  
The collision has 
also triggered a massive star formation activity in both disks as seen in 
the near and mid-infrared \citep{jarrett99}.  More recent studies of the 
Taffy I include searches for molecular gas in the bridge region.  
\citet{braine03} and \citet{gao03} have both found 
$M_{\rm H_2} = 10^9 - 10^{10} \rm M_{\odot}$ of 
molecular gas in the bridge region, which is comparable to the 
total molecular gas mass typically found in the disks of nearby spiral 
galaxies. 

The R-band image reveals that UGC~12914 is $\sim 16$ kpc wide across its 
major axis, displaying a 6 kpc long tidal arm that extends in the direction 
of the companion galaxy before eventually curling back.  Although fainter, 
a similar tidal feature is seen near the southern edge of the galaxy, 
pointing away from its companion.  The northern and southern arms  form a 
helical structure reminiscent of the galaxy system commonly known as 
the ``Tadpole Galaxy'' \citep[UGC~10214 ][]{tran03}.  
UGC~12914 harbors a bright central 
bulge, with two bright knots located $\sim 8$ kpc northwest and 
southeast from the center.   The line of sight spatial information 
is required in order to determine the exact origin of the northwestern knot; 
it is unclear whether it is associated with the northern edge of the disk 
or within the tidal arm.    At $2\micron$ \citep{jarrett99}, the brightest 
peak is found at the center of the galaxy.  The bright optical knots have 
low surface brightness infrared counterparts.   In contrast to the rather 
unusual optical characteristics seen in UGC~12914, its companion galaxy, 
UGC~12915, shows morphology and features  typical of a colliding late type 
edge-on galaxy.   A strong dust lane encircles the central region of the 
galaxy, obscuring much of the line of sight optical emission in this 
region \citep{bush90}.  In addition, a short, tidally induced warp is 
seen in the northern edge, possibly connected to the northwestern arm of 
UGC~12914.

The \HI\ distribution and kinematics, imaged using the VLA at
a $18''\times 18''$ ($5.1 \times 5.1$ kpc) resolution and shown 
in Figure~\ref{vv254}, are the same data previously published
by \citet{cond93}.
The \ion{H}{1} emission is widely spread out and  extends far beyond 
the optical disks (Figure~\ref{vv254} (\textit{top})).  
The derived total mass of neutral hydrogen is 
$1.5\times 10^{10} M_{\odot}$, $25\%$ of which 
($3.8 \times 10^9$ M$_\odot$) is found in the bridge region \citep{cond93}.  
The absence of substantial \ion{H}{1} emission in both disks may be in 
part due to absorption against the bright and extended continuum emission.  
Alternatively, a substantial fraction of \ion{H}{1} gas initially 
distributed in the main body may have already been stripped away to the 
inter-galactic medium either by ram pressure or tidal force and is   
now beginning to flow back to the disks.   

The CO~(1--0) distribution and kinematics are imaged using both 
natural weighting ($7.2''\times 5.1''$ or $2.0 \times 1.4$ kpc resolution)
and uniform weighting ($4.3''\times 3.4''$ or $1.2 \times 1.0$ kpc 
resolution), using the OVRO array. 
The CO~(1--0) emission in UGC~12914 coincides with the three optical knots 
(Figure~\ref{vv254} (see \textit{middle})).  The most massive southern
concentration has a mass of $2.4 \times 10^9 M_{\odot}$ followed by  
$1.7 \times 10^9 M_{\odot}$ and  $1.6 \times 10^9 M_{\odot}$ in the 
central and the northern concentrations.  
The southern concentration coincides with the peak in \ion{H}{1} 
emission also.  Connecting the three molecular 
complexes is a long and diffuse molecular bridge that spatially coincides 
with the main dust lane seen in the R-band image.  The total molecular gas 
mass including the diffuse bridge is $8.4 \times 10^9$M$_\odot$.  An 
elevated level of H$\alpha$ emission (Bushouse, private communication) 
gives compelling evidence of ongoing 
star formation activity near the vicinity of the three molecular complexes.   
Since the tidal arm visible in H$\alpha$ and optical images extends far 
beyond the OVRO primary beam, it is not possible to identify the presence 
of any molecular emission in the tidal tails. 

In the northern galaxy, UGC~12915, a large amount of molecular gas is 
found along the large bar-like complex along the projected optical 
major axis (Figure~\ref{vv254} (\textit{bottom})).  Three resolved CO 
peaks are identified within the main disk.  Two peaks, separated by 
5~kpc, occupy the two edges of the CO emission region.   
The southern peak is brighter of the two.   The bar-like ridge at the 
center is rather complex as it is composed of two spatially 
distinct peaks separated by less than 1 kpc.   A bright H$\alpha$ 
association is seen at the outer peaks but the central region is almost 
devoid of such emission, probably because of heavy obscuration by dust.   
The peak column density of $4.0 \times 10^{22}$~cm$^{-2}$ translates to
a mean visual extinction of $A_V \sim 80$.  This is comparable to 
that found in NGC~4647 (``The Mice''; $A_V \sim 120$) \citep[][]{yun01},
but much smaller than that found in the prototypical ULIRG Arp~220 
($A_V \sim 1000$) \citep{sakamoto99b}.  The 
estimated molecular gas mass of UGC~12915 is  $1.2 \times 10^{10}$M$_\odot$.   

There is a third major CO emitting cloud in the system, located
5~kpc southwest of UGC~12915, in the bridging region between
the two galaxies (Figure~\ref{vv254} (\textit{bottom})).
The deconvolved diameter of this isolated complex is about 2~kpc, 
and $2.7 \times 10^9$ M$_\odot$ of molecular gas (nearly
20\% of the molecular gas mass in UGC~12915) is found there.  
This bridge feature has optical, H$\alpha$, and radio counterparts, 
but it is faint in the K-band light.  
Earlier BIMA observation with a larger synthesized ($\sim 10''$) 
beam reported a larger amount ($1.4 \times 10^{10}$ M$_{\odot}$) of 
molecular gas in the bridge region \citep{gao03}.  
The association with a bright radio continuum peak suggests that 
the bridge feature is real, and a possible explanation includes a ballistic 
ejection of molecular clouds from the main body of UGC~12915 during the 
recent collision.  Similar massive ejected CO clouds have
also been found in other interacting galaxy systems such as NGC~4676
\citep[][]{yun01}. 

The \ion{H}{1} PVD (Figure~\ref{pvd1}) morphologies in both sources are 
quite irregular, probably due to the collision impact that resulted in the 
ejection of disk \ion{H}{1} out to the bridge region.  The CO~(1--0) PVD 
(Figure~\ref{pvd1}) in UGC~12914 is lopsided with strong emission arising 
from the southern concentration. The resulting CO~(1--0) rotation curve 
is consistent with that of \ion{H}{1} at $V_r \sim 300$~km~s$^{-1}$ 
and maintains a constant velocity out to a 30~kpc radius.  The high 
S/N CO~(1--0) emission in UGC~12915 results in a 
well defined PVD that traces the rotation of the disk with distinct 
emission seen in the forbidden velocity quadrant, signifying inflow 
(or outflow; see Paper~I).  The molecular gas mass associated with 
this kinematically distinct feature is $\sim 10^9$~M$_{\odot}$, 
which is less than 10\% of the total gas mass of UGC~12915.
Similar to its companion, the rotation curve derived from both \ion{H}{1} 
and CO~(1--0) are consistent at $V_r \sim 300$~km~s$^{-1}$, 
but it shows a gradual decline ($\Delta v = 100$~km~s$^{-1}$) at 
large radii (10 -- 30~kpc) probably due to a strong warp in
the gas disk (Figure~\ref{rot1}).

\subsection{VV~253 (NGC~5331, UGC~8774)}

VV~253 is one of the brightest 
(in the IR; $L_{IR} = 3.2 \times 10^{11}$~L$_\odot$) 
and the most distant galaxy pair in the 
sample. It consists of two galaxies separated by 18~kpc in the north-south 
direction, with a low surface brightness dwarf galaxy (NGC~5331W) located 
60~kpc west of the galaxy pair. The projected major axis of NGC~5331N is 
$\sim 26$ kpc with two arms extending north and south, resembling a 
reversed $\int$-sign.  The southern galaxy, NGC~5331S, is similar in size 
to NGC~5331N but appears slightly more disturbed than its companion.  The 
southern edge of NGC~5331N and the northern edge of NGC~5331S appear to 
overlap in the optical image.  

The \HI\ distribution and kinematics, imaged using the VLA at
a $22.0''\times 17.1''$ ($14 \times 11$ kpc) resolution, are shown 
in Figure~\ref{vv253}.
The peak \ion{H}{1} emission is found in the overlap region with a small 
lopsidedness toward NGC~5331N (Figure~\ref{vv253} (\textit{top})).  A 
longer but  lower column density extension covers the entire disk of 
NGC~5331S and beyond.  A secondary peak with no optical counterpart is 
found 26 kpc west of the galaxy pair, and its southwest extension connects 
to a third low column density peak that coincides with CGCG~017-081. 
Therefore the observed \HI\ distribution and kinematics suggest 
an interaction that involves at least three galaxies.  The 
\ion{H}{1} emission covers similar velocity ranges in both galaxies 
in NGC~5331, suggesting that the orbital motion is largely in the sky plane.

The distribution of the CO~(1--0) emission, imaged using the OVRO array at
a $5.4''\times 4.4''$ ($3.5 \times 2.8$ kpc) resolution
(Figure~\ref{vv253} (\textit{bottom})), shows a large 
amount ($M_{H_2} = 3.0 \times 10^{10} M_{\odot}$) of molecular gas in 
the disk of the southern galaxy NGC~5331S.  Two molecular arms are 
attached to the central concentration, and they trace the 
morphology of the optical tidal arms.  
The northern galaxy (NGC~5331N) harbors about $10\%$ of the 
molecular mass of its companion ($M_{H_2} = 3.5 \times 10^9$ M$_\odot$). 
The peak of the CO~(1--0) emission is displaced from the peak 
of the R-band image to the north by 3-4 kpc.  The K-band peak is also
similarly offset to the north by about the same amount, thereby 
suggesting strong extinction along the northern parts of NGC~5331N.  
Strong extinction toward the nucleus of NGC~5331S is also evident 
in the R- and the K-band light distribution, but spatial coincidence between 
the CO and the K-band peaks suggests a high concentration of 
molecular gas and dust near the nucleus with extended wings along the 
nearly edge-on disk. The CO velocity field shows a rotation-like 
gradient in NGC~5331S that is consistent with that of \ion{H}{1}, and
the peak of the velocity dispersion of both species coinciding with 
the central CO peak  
(see Figure~\ref{vv253} (\textit{top right}) and (\textit{bottom right})).

The \ion{H}{1} PVD (Figure~\ref{pvd1}) in NGC~5331N has an emission peak 
offset from the center of the galaxy by 15~kpc, which coincides with the 
global peak of \ion{H}{1} emission.  The low abundance of CO~(1--0) 
emitting gas in NGC~5331N (Figure~\ref{pvd1})
makes it difficult to construct a reliable CO rotation curve. 
Thus only  \ion{H}{1} emission was used to derive the rotation curve
shown in (Figure~\ref{rot1}).  The \ion{H}{1} PVD morphology 
in NGC~5331S is similar to its companion in that the same global peak is 
also included within the PVD slit.  The CO emission at the center of 
the galaxy spans the entire velocity range, with a large spread in emission 
in each velocity channel, resulting in some emission features in the 
forbidden velocity quadrants.   The rotation curve derived from these 
data show a monotonic decline from $V_r\sim 300$~\kms to 
230~\kms toward the outer disk at a radius of 23~kpc (see Figure~\ref{rot1}).  
Based on other systems studied here, this decrease in the rotation
curve probably indicates a strong warp in the outer gas disk.

\subsection{VV~247 (NGC~6621/2, UGC~11175/6, Arp~81)}

VV~247 is located at a distance of 81~Mpc with high enough infrared 
luminosity (L$_{IR} = 1.4 \times 10^{11}$ L$_{\odot}$) to be classified as 
a LIRG.  It is the fifth system in the Toomre sequence of
mergers (see \S2) with a projected separation of 17~kpc between
NGC~6621 and NGC~6622.  NGC~6621 is a late type barred spiral 
with strong $m=2$ arms extending to the northwest and southeast, where 
the northern arm curls around on the northeast side of the system 
to form a large tidal hook about 75~kpc 
in projected extent.  A large isolated star forming clump 
is seen in the overlap region 12~kpc southeast of the nucleus of NGC~6621.
The disk of NGC~6621 is 21~kpc long in the major axis and 
8~kpc in the minor axis, 
where a long extended dust lane penetrates through the nucleus and toward 
the southern isolated star forming clump.  NGC~6622 is probably an 
inclined S0 galaxy \citep{xu00} with little sign of tidal disturbance.  
It is 6~kpc long in the major axis and 4~kpc long in the minor axis.  
The high resolution HST/WFPC2 image shows a ubiquitous presence of dust lanes
in the northern and the southern edges of the galaxy system \citep{keel03}.  
Numerous star clusters were identified in the disk of NGC~6621, 
along the long tidal arm, and in the star forming clump in the 
overlap region \citep{keel03}.
Some star forming 
activity is seen in NGC~6622 as identified by its MIR and H$\alpha$ 
distribution \citep{bush90, xu00, keel03}, but at a lower level than in the 
nucleus of NGC~6621.  The star forming region seen in the overlap region 
also exhibits MIR and H$\alpha$ enhancement.  Strong FIR and 1.4 GHz radio 
emission are only detected in NGC~6621 
\citep[][Figure~\ref{radio_cont}]{cond96, bush98}, both showing a southern 
extension which is possibly related to the isolated southern star forming 
clump.  

The \HI\ distribution and kinematics, imaged using the VLA at
a $16.2''\times 13.0''$ ($6.5 \times 5.2$ kpc) resolution, 
are shown in Figure~\ref{vv247}.  NGC~6621 is one of the few galaxies 
in our sample whose disk is deficient in \ion{H}{1} (NGC~5394 is 
another example) as well as in the total 
\ion{H}{1} content ($M_{\rm HI} = 4.8 \times 10^9$ M$_{\odot}$; see 
Figure~\ref{vv247} (\textit{top})).  The \ion{H}{1} emission is centered 
around the starburst complex in the overlap region, and the 
absence of CO~(1--0) emission there makes this an interesting source for 
future research using high resolution infrared and optical imaging. 
The northern part of the tidal tail is also deficient in 
\ion{H}{1}, but a long \HI\ tail exactly tracing a similar optical 
tail is clearly seen on the northeast side of the system, 
extending over 20~kpc beyond the optical tail 
visible in the DSS image.  

The CO~(1--0) distribution and kinematics, imaged using the OVRO array at
a $6.6''\times 5.6''$ ($2.7 \times 2.3$ kpc) resolution, are shown 
in Figure~\ref{vv247}.  
Although faint in \ion{H}{1}, NGC~6621 is a bright CO source with
a total inferred molecular gas mass of $M_{\rm H_2} = 2.9 \times 10^{10}$ 
M$_\odot$ (see Figure~\ref{vv247} (\textit{bottom})).
The compact nuclear complex is the dominant CO feature in this
system, it accounts for the majority of the CO flux in this
system along with the short northwestern extension that traces the 
dust-lane in NGC~6621 and the longer extension  
to the southeast with a secondary peak located 6~kpc away. 
The southeastern extension does not reach far enough to the 
isolated star forming clump, but it curls northward to form a 5~kpc CO 
hook that has neither an obvious optical nor \ion{H}{1} counterpart.  
The CO kinematics around the nucleus show the  characteristic 
signature of rotation, but the kinematics toward the southern portion of 
the galaxy is very complex (Figure~\ref{vv247} (\textit{bottom right})).  
No CO~(1--0) emission is detected in NGC~6622 with the interferometer 
despite the reported flux of 34~Jy~\kms with the NRAO 12 meter telescope 
\citep{zhu99}.  Assuming a linewidth of 410~\kms \citep{zhu99} 
our $3\sigma$ upper limit CO line flux (6.8~Jy~\kms) in $H_2$ mass 
yields $0.5 \times 10^9$~M$_{\odot}$.
Their CO~(1--0) spectrum shows a possible line detection, 
but uncertainties in both the baseline fitting and their model 
approximations are large, and the claimed detection may not be reliable.  

The absence of significant \ion{H}{1} emission in the disk of NGC~6621 and 
the high concentration of \ion{H}{1} in the overlapping region makes the 
\ion{H}{1} PVD appear completely lopsided (Figure~\ref{pvd2}).  The shape of 
the CO~(1--0) PVD (Figure~\ref{pvd2}) appears similar to that of NGC~5331S 
due to the compact nature of the gas and a large range in velocity.  The 
wing toward the negative offset direction arises from the 
southeastern extension, 
and this results in a substantial emission in the forbidden velocity 
quadrant. The \ion{H}{1} and CO~(1--0) rotation curves of NGC~6621 are 
consistent at $\sim 250$~km~s$^{-1}$, but only a limited amount of 
information is available for the gas poor galaxy NGC~6622 (Figure~\ref{rot2}).

\subsection{VV~769 (UGC~813/6, Taffy II)}

VV~769 is investigated in detail by \citet{bush90} in optical/NIR and by 
\citet{zhu99} in CO(1--0) as part of their interacting galaxy surveys.  
\citet{cond02} identified unusual morphological and radio 
properties in UGC~813/6 that  appear similar to the bridge medium in 
Taffy~I, and thus naming it Taffy~II.  
More recent studies of the Taffy~II system include searches for 
molecular gas in the bridge region.  
Using the IRAM 30~m telescope, 
\citet{braine04} found $M_{\rm H_2} = 2 \times 10^9 \rm~M_{\odot}$ of 
molecular gas in the bridge region alone, which is comparable to the 
molecular gas mass found in the bridge region of Taffy~I (see \S8.3).
UGC~813 is nearly edge-on, and 
the south-eastern edge has a short tidal tail (4~kpc) pointing toward 
the direction of the southern edge of UGC~816.  The western edge of the 
disk appears less disturbed, but a diffuse tail is visible in high 
contrast images.  UGC~816 appears more face-on than its companion galaxy, 
and has  two tidal arms formed from the recent strong collision with 
UGC~813.


The \HI\ distribution and kinematics, imaged using the VLA at
a $16.3''\times 16.3''$ ($5.7 \times 5.7$ kpc) resolution and shown 
in Figure~\ref{vv769}, are the same data previously published
by \citet{cond02}.
The \ion{H}{1} distribution has peaks in each galaxy but is offset by a 
few kpc toward the bridge region (i.e. region between the two disks) in 
both galaxies (Figure~\ref{vv769} (\textit{top})).  The existence of a long, 
visible \ion{H}{1} tail ($\sim 20$ kpc) in UGC~816 suggests a low 
inclination disk on a prograde orbit.  However, a measurable velocity 
gradient along the tidal arm suggests a significant inclination along
the line of sight.   
The \ion{H}{1} distribution in UGC~813 appears to be less disturbed 
because its disk has a more inclined projection, but a 10~kpc long tail  
emanating from the southwest edge of the disk has a lower mean column 
density.  The strong rise in the velocity dispersion in the taffy region 
signifies the violent nature of the recent collision, similar to 
that seen in Taffy~I.

The CO~(1--0) distribution and kinematics, imaged using the OVRO array at
a $5.1''\times 4.1''$ ($1.8 \times 1.4$ kpc) resolution, are shown 
in Figure~\ref{vv769}.  
The CO~(1--0) emission in UGC~813 (Figure~\ref{vv769} (\textit{middle})) 
is distributed along the near edge-on disk, with two marginally resolved 
peaks separated by 2~kpc.  A third peak seen toward the eastern 
edge of the disk appears to be an extension of the central complex.  The 
peak of the 2MASS K-band image is offset from the CO~(1--0) complex, 
located between the western and the central CO~(1--0) complexes.  The 
slight north-south elongation of the larger and more massive northwestern 
complex results in a slightly lopsided CO~(1--0) warp.  It is also seen 
that the CO~(1--0) complexes are displaced toward the bridge region, 
consistent with the \ion{H}{1} emission.  Its companion, UGC~816 
(Figure~\ref{vv769} (\textit{bottom})), harbors two comparable mass 
(M$_{H_2}$  = $1.0 \times 10^9$ M$_\odot$) molecular gas complexes 
separated by 5~kpc that is further connected by a less massive 
(M$_{H_2}$  = $0.5 \times 10^9$ M$_\odot$) molecular bridge.  The two 
large CO~(1--0) complexes are displaced by $\sim 2$ kpc toward the leading 
edge of the southern spiral arm with the bridge displaced roughly twice as 
much.  Similar to UGC~813 the CO~(1--0) displacement is consistent with 
the \ion{H}{1} displacement.  

The \ion{H}{1} PVDs (Figure~\ref{pvd2}) in both UGC~813 and UGC~816 have 
similar characteristics to that of UGC~12914/5.   The emission region 
generally slopes from the upper left to the lower right quadrant, but  
well defined rotation is absent.  This is primarily due to the widely 
distributed and highly disturbed nature of the \ion{H}{1} 
emitting clouds in this recent collision system. 
The CO~(1--0) PVD (Figure~\ref{pvd2}) in UGC~813 
appears to have significant emission in the forbidden velocity quadrant, 
but the clumpy nature of the emission makes the determination of the 
systemic velocity very uncertain.
The \ion{H}{1} and CO~(1--0) rotation curves (Figure~\ref{rot2}) 
in UGC~813 are inconsistent and they differ by $\sim 100$~km~s$^{-1}$, and the 
outer disk shows a decline in \ion{H}{1} rotation similar to that seen in 
NGC~5257.  The clumpy and asymmetric distribution of CO emission 
in UGC~816 introduces significant 
uncertainties in the derivation of the CO rotation curve and 
results in a significant deviation from the \ion{H}{1} rotation curve.

\subsection{VV~242 (UGC~11984/5, NGC~7253, ARP~278)}

VV~242 is located at a distance of 57 Mpc and consists of two highly 
inclined galaxies separated by a projected distance of 12 kpc.    
UGC~11984 is 25~kpc long  with the northwestern edge harboring a 
9~kpc long low surface brightness, lopsided warp, tilted by 45 degrees 
from the disk.  This is connected to a lower surface brightness 
(possibly tidal) hook that curls toward the southwestern edge of UGC~11985.  
The southeastern edge of UGC~11984  appears to overlap with the nucleus 
of UGC~11985, where a NIR peak and an extended H$\alpha$ emission is 
found \citep{xu00}.  Although slightly shorter, the length of the projected 
major axis of UGC~11985 is similar to its companion ($\sim 20$ kpc), also 
showing similar tidal features in the northwestern edge.  Several H$\alpha$ 
knots are detected in both disks, with extended NIR and radio continuum 
emission peaking at the nucleus of UGC~11984 \citep{cond96, xu00}.  Both 
disks show evidence of nuclear starbursts where a model fit to the spectra 
suggest a $1 - 7 \times 10^7$ year time delay in the onset of the nuclear 
starburst activity \citep{bernlohr93}.

The \HI\ distribution and kinematics, imaged using the VLA at
a $17.9''\times 17.6''$ ($5.5 \times 5.4$ kpc) resolution, 
are shown in Figure~\ref{vv242}.  The
\ion{H}{1} emission, 46 kpc wide, envelops both galaxies with 
a depression of \ion{H}{1} toward the nucleus of UGC~11984 
(see Figure~\ref{vv242} (\textit{top})).  
Despite optical evidence of tidally induced tails in the 
northwestern edge of UGC~11984 pointing south, the morphology of the low 
column density \ion{H}{1} contours shows an extension to the north.  The 
existence of the northern \ion{H}{1} extension is puzzling since a general 
tendency of galaxy collisions is to induce tidal features in the plane of the 
disk, unless a direct head-on collision results in the collisions 
of gas clouds at the impact, pulling material out of the plane of the disks, 
much like the bridge material seen in Taffy I.   
A similar feature is seen in M82, where the direct ``pole-on'' passage
of M81 caused a strong double warp \citep{yun99}.
Much of the \ion{H}{1} emission in 
UGC~11985 is confined to the disk with almost all of the northern part 
attached to the \ion{H}{1} in its companion.  Since some of the \ion{H}{1} 
emission show  physical coupling with the emission in the southeastern edge 
of UGC~11985, it is likely that the two galaxies are in close physical 
contact.  However, the possibility that the line of sight projection 
mimics the apparent physical coupling cannot be ruled out.  

The CO~(1--0) distribution and kinematics, imaged using the OVRO array at
a $4.4''\times 3.1''$ ($1.3 \times 0.9$ kpc) resolution, are shown 
in Figure~\ref{vv242}.  
Despite the evidence for nuclear starburst and tidally disrupted \ion{H}{1}, 
the distribution and kinematics of the CO~(1--0) emission in 
UGC~11984 (Figure~\ref{vv242} (\textit{bottom})) appear to be relatively 
unaffected by the tidal activity, resembling a normal edge-on spiral 
galaxy.  The peak CO~(1--0) 
emission is found at the nucleus of UGC~11984 with emission extending 11~kpc 
long along the edge-on disk, where the southeastern side shows longer and 
narrower emission than the northwestern edge.  Its kinematics show a 
northwest-southeast gradient with a steeper gradient on the northwestern 
side.  The velocity dispersion peaks roughly near the K-band nucleus.  
No CO~(1--0) emission was detected in UGC~11985.  
Assuming a linewidth of 263~km~s$^{-1}$ \citep{zhu99} 
our $3\sigma$ upper limit of H$_2$ mass in UGC~11985 
yields $3.3 \times 10^8$~M$_{\odot}$.  

Low abundance of \ion{H}{1} emitting clouds near the center of 
UGC~11984 results in a \ion{H}{1} PVD with two distinct peaks  
(Figure~\ref{pvd2}), but the higher density 
tracer in CO~(1--0) emission fills the inner region.  The CO~(1--0) PVD 
(Figure~\ref{pvd2}) shows  a well defined rotation with the peak slightly 
offset toward the lower right quadrant.  The CO and  the 
\ion{H}{1} rotation curves derived from these PVDs are consistent with
$V_r \sim 220$~km~s$^{-1}$ (Figure~\ref{rot2}).   
The \ion{H}{1} PVD in UGC~11985 is similarly 
lopsided, and the derived rotation curve shows a gradual decline 
($\Delta v \sim 50$~km~s$^{-1}$) toward large radii (5 -- 10~kpc)
-- a possible signature for a tidally driven warp.

\subsection{VV~219 (NGC~4567/8, UGC~7776/7)}

Two early type (Sbc) spiral galaxies constitute the interacting 
Virgo cluster system 
VV~219.  The absence of any obvious large scale optical tidal features 
in either system, the proximity of the two disks with projected nuclear 
separation of $\sim 6$ kpc, and similar radial velocities 
($v_{sys}$ = 2274 (NGC~4567) and 2255 (NGC~4568)) give compelling evidence 
of a young interacting system.  
NGC~4567 is a low inclination early-type spiral galaxy 
($10 \times 5$ kpc) with two arms from the northern and southern side of 
the disk.    The northern arm curls toward the northern edge of its 
companion where dust lanes are visible at the contact region.  The southern 
arm extends in the opposite direction.  NGC~4568 is $14 \times 5$ kpc in 
size and its undisturbed appearance resembles a normal early-type spiral 
galaxy when viewed in the absence of its companion.  Numerous dust lanes 
and strong extended H$\alpha$ emission are  seen throughout the disk, 
where the most intense emission is found within $r \leq 4$ kpc in 
NGC~4568 \citep{koopmann01}.  The radio continuum centers around the 
nucleus of NGC~4568 with an extension toward its companion 
(\citet{cond96}, Figure~19).

The \HI\ distribution and kinematics, imaged using the VLA at
a $19.9''\times 14.0''$ ($1.6 \times 1.1$ kpc) resolution, 
are shown in Figure~\ref{vv219}. 
The peak  \ion{H}{1} emission is located on the northern edge of NGC~4568 
where the two galaxies appear to overlap (see Figure~\ref{vv219} 
(\textit{top})).  The emission peaks broadly trace the optical
spiral arms while the \ion{H}{1} emission has an obvious hole
in the central regions of both NGC~4567 and NGC~4568.  Such a
distribution is commonly seen in nearby spiral galaxies 
\citep[e.g. ][]{bosma81} and is remarkably unexceptional.  
The extent of the \HI\ disks
are comparable to the stellar disks in both galaxies.  The outer
\HI\ contours show a warp-like feature along the southern tip of
NGC~4568, but there is little else that suggests any recent tidal
disruptions around either of the galaxies. 
Because these galaxies reside in a cluster environment, their outer 
disks are expected to be deficient in \ion{H}{1} from repeated ram 
pressure stripping by the intercluster medium
\citep{cayatte90}.  Any loosely bound tidal material may have
also been swept away by the ram pressure stripping.  
This should occur more intensively near the cluster core as 
evidenced by the progressively smaller $D_{HI}/D_{25}$ ratio seen
among the spiral galaxies toward the core \citep{cayatte94}.  
VV~219 is located at only 650 kpc away in projected distance
from the cluster center, and NGC~4567 is one of the \HI\ anemic
galaxies found in the cluster.

The CO~(1--0) distribution and kinematics, imaged using the OVRO array at
a $4.0''\times 3.5''$ ($0.3 \times 0.3$ kpc) resolution, are shown 
in Figure~\ref{vv219}.  
The clumpy CO~(1--0) distribution in NGC~4567 is composed of a 
central molecular complex and the northern and the southern 
extension that appears to trace optical arms and dust lanes 
(see Figure~\ref{vv219} (\textit{middle})).  NGC~4568 appears to harbor a 
central molecular bar a few kpc in extent, connected with two $m=2$ 
extension stretching northeast and southwest, together forming a flipped 
$\int$-sign (see Figure~\ref{vv219} (\textit{bottom})).  At larger scales, 
molecular gas appears to trace the spiral arms that are the most
prominent structures in the outer disk.  
The ``butterfly'' shape of the velocity 
distribution seen in both \ion{H}{1} and CO is typically observed 
in rotationally supported isolated spiral galaxies.  Together with the 
lack of evidence for obvious tidal features in optical and in \ion{H}{1}, 
this may suggest that VV~219 is now commencing its initial collision that 
marks the epoch  $t \sim 0$ in the model analysis in Paper~I.  

The \ion{H}{1} PVD of NGC~4567 (Figure~\ref{pvd2}) 
is lopsided toward the lower left 
quadrant because the \ion{H}{1} emission from the northern arm of NGC~4568 
is included in the PVD slit.  Thus the \ion{H}{1} rotation curve at large 
radii (beyond $\sim 0.7'$) should be neglected.  The CO~(1--0) PVD 
of NGC~4567 (Figure~\ref{pvd2}) is similarly lopsided but at a 
much smaller scale 
($\sim 5''$) than in \ion{H}{1}.  Overall, the \ion{H}{1} and CO 
rotation curves are consistent in velocity at $\sim 80$~km~s$^{-1}$, 
but the \ion{H}{1} rotation velocity rises gradually to 120~km~s$^{-1}$ 
in the radius range of 2 -- 4~kpc (see Figure~\ref{rot2}).  
The \ion{H}{1} PVD in NGC~4568 has two bright emission features both offset 
by $\sim 5$~kpc from the center of the galaxy.  The central region of the  
\ion{H}{1} PVD is deficient in \ion{H}{1} emission, 
and this is where the CO~(1--0) emission dominates.  The CO 
PVD in NGC~4568 peaks along the rising part of the rotation with 
a slight depression of CO~(1--0) near the K-band nucleus, 
and this may suggest the presence of a 
nuclear ring.  The rotation curves derived from both of these  PVDs are 
relatively well constrained in velocity at $V_r \sim 150$~\kms.

\subsection{VV~731 (NGC~7592,  MRK~928)}

The optical image of the luminous infrared galaxy, VV~731, 
suggests that this might possibly be a triplet system. 
Two large galaxies (NGC~7592E and NGC~7592W) 
separated by 7~kpc in the east-west direction and a smaller possible 
companion (NGC~7592S) located directly south of NGC~7592E form the vertices 
of a small equal lateral triangle.  This morphology and the
projected separation of 7 kpc between the two nuclei has led 
to the inclusion of this pair as the 3rd object in the Toomre 
Sequence of mergers (see \S2).  
NGC~7592W is classified as a Seyfert 2 and NGC~7592E is found to be an 
\ion{H}{2} galaxy \citep{veron97, rafanelli92}.  Two long tidal tails 
both 15~kpc in projected length emanates from the northern side of 
NGC~7592W and the southeastern side of NGC~7592E.  No MIR emission is 
detected from the small companion \citep{hwang99}.   The high resolution 
HST WFPC2 image \citep{malkan98} and ground based H$\alpha$ imaging 
\citep{dopita02} both suggest that NGC~7592S is a large massive star 
forming region  in the tidal tail of NGC~7592E, similar to what is seen 
in the southern clump of NGC~6621. The HST/WFPC2 image also reveals that 
NGC~7592E is more disturbed than its companion, with numerous dust-lanes 
encircling and penetrating the nuclear region.  \citet{rafanelli92} found 
a large star formation rate ($\sim 20 M_\odot$ yr$^{-1}$) from their 
emission line analysis and the presence of ionized gas at the interface 
between NGC~7592W and NGC~7592E.  

The \HI\ distribution and kinematics, imaged at
a $129''\times 83''$ ($36 \times 23$ kpc) resolution
using the archival VLA data obtained in the D-array in 1983, 
are shown in Figure~\ref{vv731}.  The low angular resolution
and the poor sensitivity resulting from a short duration of the
observations makes a detailed investigation of 
the emission morphology difficult.  In contrast to the other program sources 
mapped in this study, the peak \ion{H}{1} emission is offset from the two 
galaxies by a large amount ($\sim 2.5' = 71$~kpc; see Figure~\ref{vv731} 
(\textit{top})).  The exact coincidence of the radio continuum peak
with the stellar systems (Figure~\ref{radio_cont}) suggests that the 
astrometry is correct and the \HI\ displacement real.  
Such a large offset of \ion{H}{1} emission is typically seen in 
late stage mergers \citep{hibbard96,hibyun96}.  
The velocity gradient in the \HI\ tail is in the north-south direction
(Figure~\ref{vv731}~(\textit{top})),
similar to the sense of rotation implied by the CO~(1--0) 
velocity gradient (Figure~\ref{vv731}~(\textit{bottom})).
This may indicate that the long \HI\ tail originates from the 
disk of NGC~7592W, and future observations
with higher resolution and sensitivity will be able to address 
the \HI\ distribution and kinematics much better.

The CO~(1--0) distribution and kinematics, imaged using the OVRO array at
a $4.5''\times 3.6''$ ($2.2 \times 1.7$ kpc) resolution, are shown 
in Figure~\ref{vv731}.  
The peak  of CO~(1--0) emission coincides with the two optical nuclei 
(see Figure~\ref{vv731} (\textit{bottom})), each with a nearly identical
molecular gas mass of $7\times 10^9 M_\odot$.  A low column density extension 
($M_{H_2} = 7 \times 10^8$ M$_\odot$) connects the two galaxies, 
and this discovery is consistent with the detection of ionized gas 
in the same region \citep[][]{rafanelli92}.  The CO emission in NGC~7592W 
is elongated from northwest to southeast, but the velocity 
gradient runs almost perpendicular to it.  Such an unusual velocity
field is sometimes seen in face-on systems (e.g. NGC~5394 in
Fig.~\ref{vv48} (\textit{middle})).  The largest velocity
dispersion in NGC~7592W is found near the southeastern part of the 
galaxy.  Along with the unusual velocity gradient, this is evidence for 
non-circular kinematics caused by the recent interaction.  
The distribution of CO~(1--0) emission in NGC~7592E is elongated 
east-west with a velocity gradient 
in the same direction.   The molecular gas mass for the CO emission
in NGC~7592S ($M_{H_2} = 6 \times 10^8$ M$_\odot$) is about 10\% of 
that detected in NGC~7592E.  


The \ion{H}{1} PVD is not constructed because of the poor 
quality of the data.  The CO PVDs (Figure~\ref{pvd3}) for 
two galaxies display peaks in the lower left quadrants, with secondary 
peaks near the tip of the rising part of the rotation in the upper right 
quadrants.  Some emission in the forbidden velocity quadrants are detected  
in both galaxies.  Because of the lack of velocity information on larger 
scales, the CO rotation curves derived from these PVDs offer 
limited information (Figure~\ref{rot3}).   The rotation velocity of 
these two galaxies are comparable at $V_r \sim 200$~km~s$^{-1}$.

\subsection{VV~244 (NGC~5953/4, UGC~9903/4, ARP~91)}

At a distance of 24~Mpc, VV~244, is one of the nearest 
galaxy pairs in our sample 
with a projected nuclear separation of just 6~kpc.   NGC~5953 is a face-on 
spiral 6 kpc across with a large bulge dominating the R-band light
distribution.    NGC~5953 is classified as a LINER by 
\citet{veilleux95}, and as a Seyfert 2 by \citet{gonzalez96}.  A 
burst of circumnuclear star formation may have been induced by the 
interaction \citep{gonzalez96}.  Soft X-ray emission has been detected
in NGC~5953, probably related to the Seyfert activity based on the
observed variability \citep{pfefferkorn01}.   Its companion, NGC~5954,  
is 8~kpc long in the projected major axis and 4~kpc long in the minor 
axis.   An optical image shows a southern tail that curls northward 
forming a small tidal hook.  It harbors a LINER nucleus \citep{gonzalez97} 
with no signs of Seyfert activity \citep{pfefferkorn01}.  The 
HST/WFPC2 F606W imaging by \citet{malkan98} reveals 
widespread dust lanes and the southern tidal hook in NGC~5953 as well as 
the numerous dust lanes that occupy the central region in NGC~5954.
Interferometric \ion{H}{1} observations of VV~244 were previously 
reported by \citet{chengalur95}, who 
found a long \ion{H}{1} plume extending to the northwest of the pair where 
only a faint,  diffuse optical counterpart is known to exist. They suggested 
that the plume arises mostly from the tidally ejected gas from NGC~5954, 
with a small contribution from NGC~5953.  From their N-body experiments to 
model the optical morphology seen in NGC~5954, \citet{jenkins84} derived 
an interaction age of 40~Myr, nearly independent of the orbital parameters.  

The \HI\ distribution and kinematics, imaged using the VLA at
a $19.8''\times 17.9''$ ($2.3 \times 2.1$ kpc) resolution, 
are shown in Figure~\ref{vv244}. 
The large scale features such as the long \HI\ plume reported by 
\citet{chengalur95} are clearly seen.  In addition, the superior 
surface brightness sensitivity of our new data reveal
fainter \HI\ features such as the extension to the northeast with a 
projected distance of $\sim 20$ kpc from the mean position of the galaxy 
pair (see Figure~\ref{vv244} (\textit{top})).  A separate \ion{H}{1} hook 
originates from the northwestern side of NGC~5953 extending northward 
by $\sim 6$ kpc.  It merges with the diffuse emission connected 
to the northern side of NGC~5954, together forming the long \ion{H}{1} 
plume.  The \ion{H}{1} emission peaks on both galaxies are systematically 
shifted toward the faint optical bridge region, as seen in many of
the early collision systems studied here.  

The CO~(1--0) distribution and kinematics, imaged using the OVRO array at
a $4.4''\times 3.6''$ ($0.6 \times 0.5$ kpc) resolution, are shown 
in Figure~\ref{vv244}.   The  molecular gas distribution in NGC~5953 
is symmetric with a short extension 
pointing toward NGC~5954 (see Figure~\ref{vv244} (\textit{middle})).  The 
deconvolved size of the CO emitting region is $1.8 \times 1.4$ kpc, and 
it fills about 20\% of the R-band disk.   On the other hand, the CO~(1--0) 
emission in NGC~5954 (Figure~\ref{vv244} (\textit{bottom})) is more  
densely concentrated near the nucleus with an asymmetric southern 
extension along the more dominant western stellar arm.  The northern and 
western side of the galaxy are almost devoid of molecular gas.   The 
velocity field in NGC~5953 appears to follow a normal circular rotation 
where the velocity increases from the northeast to the southwest.   The 
velocity field in NGC~5954 exhibits a north-south gradient, which is 
consistent with the sense of rotation inferred from the optical morphology.  

The \ion{H}{1} PVD in NGC~5953 (Figure~\ref{pvd3}) is significantly 
lopsided toward the upper left quadrant, with  low velocity features 
covering a larger extent ($\sim 7$~kpc). The CO PVD in NGC~5953  
(Figure~\ref{pvd3}) seemingly traces the rotation well, and the 
emission peaks are located near the tip of the rotation.  The CO 
rotation curve declines monotonically from $V_r \sim 130$~km~s$^{-1}$ 
to 80~km~s$^{-1}$ and does not connect smoothly to 
that of the \ion{H}{1} rotation curve, which has $V_r \sim 150$
km~s$^{-1}$ and rises to 200~km~s$^{-1}$ at larger radii ($1'' = 7$~kpc). 
The \ion{H}{1} PVD in NGC~5954 is similarly lopsided 
toward the lower right quadrant, but the CO PVD appears to trace 
a circular rotation.  Similar to NGC~5953, the \ion{H}{1} and CO 
rotation curves do not agree exactly and are  
offset by $\Delta v = 50$~\kms (Figure~\ref{rot3}), 
possibly due to tidal stripping of the disk \HI\ gas.

\section{Summary and Discussions}  

Atomic and molecular gases are the dominant constituents of the ISM in 
galaxies, and tidal perturbations from a companion of comparable mass can 
significantly alter the distribution and kinematics, leading to
compression, inflow, and enhanced star formation activities
\citep[e.g. ][]{mihos96}.  As an observational test of such a 
theoretical scenario, we have identified a sample of 10 pairs of
nearly equal mass large disk galaxies in their early stages of
interaction or merger and obtained new or archival  
\ion{H}{1} and CO~(1--0) data using the VLA and the OVRO
millimeter array for the entire sample.  
The sensitivity and angular resolution of our data are sufficient to
allow detailed analysis of \HI\ and CO distribution and kinematics
in the complete sample.
Some of the key analysis conducted in this work are summarized below.

\begin{enumerate}
\item \textbf{Atomic and molecular gas masses:}   The total neutral
atomic gas mass ranges from $1.8 \times 10^9 M_\odot$ to 
$3.4 \times 10^{10} M_\odot$.  
The total molecular gas mass derived from the CO luminosity
ranges from $7.4 \times 10^8 M_\odot$ to 
$4.5 \times 10^{10} M_\odot$.  More than half of the sources have 
higher molecular gas mass than atomic gas, and this may suggest the 
dominance of molecular gas in these colliding galaxies.

\item \textbf{Distribution of the \ion{H}{1} emission:}
\ion{H}{1} emission in 50\% of the sample sources display long \ion{H}{1} 
tails, indicative of at least one of the galaxies in the pair following  
a prograde orbit.  
Four out of five (80\%) systems with long \HI\ tails are 
classified as LIRGs, 
suggesting a correlation between tidal disruption of
outer gas and disk star formation activity. 
The $10^{21}$~cm$^{-2}$ column density contours 
in nearly 80\% of the sources are connected between the two galaxies 
involved in the collision (i.e. Taffy like sources).  The environment 
does not appear to have a significant impact on the development of these 
large \ion{H}{1} tidal features.  The \ion{H}{1} peaks are offset
from the stellar light distribution in six of the groups (VV~244 VV~247, 
VV~253, VV~254, VV~731, and VV~769; 60\%).

\item \textbf{Distribution of the CO~(1--0) emission:} 
The CO~(1--0) emission peak in four galaxies 
(NGC~5258, NGC~5331N, NGC~5395,  and 
UGC~12914; i.e. $\sim 20$\%) is offset from the dynamical 
center of the galaxy determined from the 2MASS K-band images.  There are five 
CO~(1--0) emission complexes with no optical counterpart (UGC~12914, 
UGC~816, NGC~5331S, NGC~5954 and NGC~6621; i.e. $\sim 25$\% of the galaxies). 
 
\item \textbf{Position Velocity Diagrams:} More than 50\% of the 
sources in CO~(1--0) shows a central hole at the PVD centroid, and this 
is indicative of an existence of a central gas ring.  At least 10\% of the 
sources show substantial CO~(1--0) emission in the 
forbidden velocity quadrant which is a possible evidence of radial inflow 
(Paper~I).

\item \textbf{Rotation Curves:} 
Out of the 15 galaxies with robust derivation of the \ion{H}{1} and 
CO~(1--0) rotation curves, five galaxies ($\sim33$\%; NGC~5395, UGC~813, 
UGC~816, NGC~5953, NGC~5954) shows significant departure 
($\Delta v > 50$~km~s$^{-1}$) between the \ion{H}{1} and CO rotation 
curves.  This may be explained by the fact that molecular gas in general 
has a clumpy distribution with a lower filling factor than \HI, 
and that the loosely bound disk \HI\ gas is tidally ejected.

\end{enumerate}

The derived physical properties, distribution, and kinematics 
of the gas were presented in this paper in order to build a statistically 
significant sample of gas in colliding systems. The characteristics of the 
physical parameters and the structural parameters can potentially 
systematically diverge from the same parameters derived for isolated spiral 
galaxies in the local universe. 
These analysis using the BIMA SONG data are detailed in Paper~III.

D.I. would like to thank his thesis committee members both at UMass and CfA,
Neal Katz, James Lowenthal, Edward Chang, David Wilner, Gary Melnick 
and Mark Reid, and Qizhou Zhang, 
and his fellow students, Junzhi Wang, Peter Sollins,
Dusan Keres, Jun-Hwan Choi, Jenny Greene for fruitful discussion. He also 
thanks James Battat for careful editing of the thesis manuscript.  
The authors would like to thank the anonymous referee for the valuable
and prompt feedback.

The work presented here is part of D.I.'s Ph.D. thesis at the 
University of Massachusetts, and the bulk of the work was done during
D.I.'s predoc fellowship at the Smithsonian Astrophysical Observatory.
This research is partly supported by the Faculty
Research Grant at the University of Massachusetts and the National 
Science Foundation grant AST-0228993.

\appendix

\section{Spin Temperature from a Three Cloud Model}
The detected \ion{H}{1} emission from a narrow velocity channel can 
be comprised of blended emission from several overlapping clouds with 
different temperature and density.  In order to model the emerging 
temperature from the blended emission, 
consider the simple 3 independent thermalized cloud model in 
Figure~\ref{spintemp_fig1}.  Here cloud 1 and cloud 3 have exactly 
the same properties and they represent the warm neutral medium (WNM) 
and cloud 2 represents the cold medium (CM) of the ISM.  This models a 
simple \ion{H}{1} cloud structure in a disk, and it is motivated by the 
observed evidence that the scale height of the WNM is about twice as 
large as that of the CM in the Galaxy.  Therefore, assuming a face on 
galaxy, the emission from the far side of the galaxy (cloud 3) goes 
through the CM (cloud 2), then through the WNM (cloud 1) and arrives 
at the observer.  The emerging brightness temperature at the observer 
can be expressed as,
\begin{equation}
T_b = T_1 \left(1-e^{-\tau_1}\right) + \left[ T_2 \left( 1-e^{-\tau_2} \right) + T_3 \left( 1-e^{-\tau_3} \right) e^{-\tau_2} \right] e^{-\tau_1}.
\end{equation}
The blended brightness temperature can then be converted to spin 
temperature assuming $\tau \simeq \tau_1 + \tau_2 + \tau_3$.  The 
optical depths ($\tau_1 = \tau_3$ and $\tau_2$) are both related to 
the temperature and density (i.e. $\tau \propto N/T$).  Therefore, 
Equation~A1 can be solved easily once an assumption for the density 
of the clouds are made.  Figure~\ref{spintemp_fig} shows the results 
for two cases; $N_1 = N_3 = 0.1N_2$ and $N_1 = N_3 = 0.5N_2$.

\begin{figure}
  \caption{Schematic of the two temperature three cloud model.}
\label{spintemp_fig1}
\end{figure}

\begin{figure}
  \caption{Spin temperature as a function of the optical depth of 
cloud 2.  This experiment demonstrates that the spin temperature of 
the ISM is of order 100~K at low optical depths (i.e. $\tau << 1$).}
\label{spintemp_fig}
\end{figure}

The results demonstrated that the temperature of the ISM is of order 
100~K at low optical depths ($\tau << 1$) regardless of the cloud 
densities.  
When the relative density of the WNM is high (i.e. $N_1 = N_3 = 0.5N_2$), 
the spin temperature increases more rapidly as a function of $\tau_2$ 
because the contribution from cloud 1 becomes more significant at higher 
optical depths.  In contrast, when the relative density of the WNM is low 
(i.e. $N_1 = N_3 = 0.1N_2$), then the dominant contribution arises from 
cloud 2 regardless of the optical depth, and therefore a shallower slope.


\begin{thebibliography}{fun}
\bibitem[Abazajian et al.(2003)]{abazajian03} Abazajian et al. 2003, AJ, 126, 2081
\bibitem[Bernl\"ohr(1993)]{bernlohr93} Bernl\"ohr, K. 1992, A\&A, 268, 25
\bibitem[Blain et al.(2002)]{blain02} Blain, A. W., Smail, I., Ivison, R. J., Kneib, J. P., \& Frayer, D. T.,  2002, PhR, 369, 111
\bibitem[Bosma(1981)]{bosma81} Bosma, A. 1981, AJ, 86, 1791 
\bibitem[Braine et al.(2003)]{braine03} Braine, J., Davoust, E., Zhu, M., Lisenfeld, U., Motch, C., \& Seaquist, E. R. 2003, A\&A, 408, 13 
\bibitem[Braine et al.(2004)]{braine04} Braine, J., Lisenfeld, U., 
Duc, P. -A., Brinks, E., Charmandaris, V., \& Leon, S. 2004, A\&A, 418, 419
\bibitem[Bryant \& Scoville(1999)]{bryant99} Bryant, P. M., \& Scoville,
N. Z. 1999, AJ, 117, 2632
\bibitem[Bushouse(1986)]{bushouse86} Bushouse, H. A. 1986, AJ, 91, 255 
\bibitem[Bushouse(1987)]{bushouse87} Bushouse, H. A. 1987, ApJ, 320, 49 
\bibitem[Bushouse \& Werner(1990)]{bush90} Bushouse, H. A., \& 
Werner, M. W.  1990, ApJ, 359, 72
\bibitem[Bushouse, Telesco \& Werner(1998)]{bush98} Bushouse, H. A., 
Telesco, C. M. \& Werner, M. W.  1998, AJ, 115, 938
\bibitem[Cayatte et al.(1990)]{cayatte90} Cayatte, V., van Gorkom, 
J. H., Balkowski, C. \& Kotanyi, C. 1990, AJ, 100, 604
\bibitem[Cayatte et al.(1994)]{cayatte94} Cayatte, V., Kotanyi, C., 
Balkowski, C. \& van Gorkom, J. H. 1994, AJ, 107, 1003
\bibitem[Chengalur, Salpeter \& Terzian(1995)]{chengalur95}  Chengalur, 
J. N., Salpeter, E. E. \& Terzian, Y. 1995, AJ, 110, 167
\bibitem[Cole et al.(2000)]{cole00} Cole, S., Lacey, C. G., Baugh, 
C. M., \& Frenk, C. S. 2000, MNRAS, 319, 168
\bibitem[Condon et al.(1993)]{cond93} Condon, J. J., Helou, G., Sanders, 
D. B., \& Soifer, B. T.  1993, AJ, 105, 1730 
\bibitem[Condon et al.(1996)]{cond96} Condon, J. J., Helou, G., Sanders, 
D. B., \& Soifer, B. T.  1996, ApJS, 103, 81
\bibitem[Condon et al.(1998)]{cond98} Condon, J. J., Cotton, W. D., 
Greisen, E. W., Yin, Q. F., Perley, R. A., Taylor, G. B., \& Broderick, 
J. J. 1998, AJ, 115, 1693
\bibitem[Condon, Helou \& Jarrett(2002)]{cond02} Condon, J. J., Helou, 
G., \& Jarrett, T.H.  2002, AJ, 123, 1881 
\bibitem[Dopita et al.(2002)]{dopita02} Dopita, M. A., Pereira, M., 
Kewley, L. J \& Capaccioli, M. 2002, ApJS, 143, 47
\bibitem[Downes \& Solomon(1998)]{downes98} Downes, D. \& Solomon, 
P. M. 1998, 507, 615
\bibitem[Gao \& Solomon(1999)]{gao99} Gao, Y. \& Solomon, P. M. 1999, 
ApJ, 512, 99
\bibitem[Gao, Zhu \& Seaquist(2003)]{gao03} Gao, Y., Zhu, M., \& 
Seaquist, E. R. 2003, AJ, 126, 217
\bibitem[Georgakakis, Forbes \& Norris(2000)]{georgakakis00} 
Georgakakis, A., Forbes, D. A. \& Norris, R. P. 2000, MNRAS, 318, 124
\bibitem[Gonzalez-Delgado \& Perez(1996)]{gonzalez96} Gonzalez-Delgado, 
R. M. \& Perez, 1996, MNRAS, 281, 781 
\bibitem[Gonzalez-Delgado et al.(1997)]{gonzalez97}  Gonzalez-Delgado, 
R. M., Perez, E., Tadhunter, C., Vilchez, J. M. \& Rodriguez-Espinosa, 
J. M.,  1997, ApJS, 108, 155
\bibitem[Gooch(1995)]{gooch95} Gooch, R.E., 1995, "Space and the 
Spaceball", in Astronomical Data Analysis Software and Systems IV, 
ASP Conf. Series vol. 77, ed. R.A. Shaw, H.E. Payne, \& J.J.E. Hayes, 
ASP, San Francisco, p.144-147, ISBN 0-937707-96-1
\bibitem[Helou et al.(1988)]{helou88} Helou, G., Khan, I. R., Malek, L.,
\& Boehmer, L. 1988, ApJS, 68, 151
\bibitem[Helfer et al.(2002)]{helfer02} Helfer, T. T., Vogel, S. N., 
Lugten, J. B. \& Teuben, P. J. 2002, PASP, 114, 350
\bibitem[Helfer et al.(2003)]{helfer03} Helfer, T. T., Thornley, M. D., 
Regan, M. W., Wong, T., Sheth, K., Vogel, S. N., Blitz, L \& Bock, 
D. C.-J. 2003, ApJS, 145, 259
\bibitem[Hernquist \& Mihos(1995)]{hernquist95} Hernquist, L. \& Mihos, 
J. C. 1995, ApJ, 448, 41
\bibitem[Hibbard \& van Gorkom(1996)]{hibbard96} Hibbard, J. E. \& 
van Gorkom, J. H.  1996, AJ, 111, 655
\bibitem[Hibbard \& Yun(1996)]{hibyun96} Hibbard, J. E., \& Yun, M. S.
1996, in {\it Cold Gas at High Redshift,} eds. M.N. Bremer \& N. Malcolm,
Astrophysics and Space Science Library, Vol. 206, p.47 (Dordrecht: Kluwer Academic Publishers)
\bibitem[Horellou, Booth \& Karlsson(1999)]{horellou99} Horellou, C., 
Booth, R. S. \& Karlsson, B. 1999, Ap\&SS, 269, 629
\bibitem[Hwang et al.(1999)]{hwang99} Hwang, C. -Y., Lo, K. Y., Gao, 
Y., Gruendl, R. A. \& Lu, N. Y. 1999, ApJL, 511, 17
\bibitem[Iono, Yun \& Mihos(2004)]{iono04} Iono, D. Yun, M. S. 
\& Mihos, C. J. 2004, ApJ, 616, 199 (Paper~I)
\bibitem[Jarrett et al.(1999)]{jarrett99} Jarrett, T. H., Helou, G., 
Van Buren, D., Valjavec, E., \&  Condon, J. J. 1999, AJ, 118, 2132
\bibitem[Jarrett et al.(2003)]{jarrett03} Jarrett, T. H., Chester, 
T., Cutri, R., Schneider, S. E. \& Huchra, J. P.  2003, AJ, 125, 525
\bibitem[Jenkins(1984)]{jenkins84}  Jenkins, C. R.  1984, ApJ, 277, 501
\bibitem[Kaufman et al.(1999)]{kaufman99}  Kaufman, M., Sheth, K., 
Struck, C., Elmegreen, B. G., Thomasson, M., Elmegreen, D. M. \& Brinks, 
E., 1999, AJ, 123, 702 
\bibitem[Kaufman et al.(2002)]{kaufman02}  Kaufman, M., Brinks, E., 
Elmegreen, B. G., Elmegreen, D. M., Klaric, M., Struck, C., 
Thomasson, M., \& Vogel, S., 2002, AJ, 118, 1577 
\bibitem[Keel \& Borne(2003)]{keel03} Keel W. C. \& Borne K. D. 
2003, AJ, 126, 1257 
\bibitem[Kenney et al.(1992)]{kenney92} Kenney, J. D. P., Wilson, C., D., 
Scoville, N., Z., Devereux, N., A., \& Young, J., S. 1992, ApJ, 395, 79L
\bibitem[Koopmann, Kenney \& Young(2001)]{koopmann01} Koopmann, R. A., 
Kenney, J. D. \& Young, J. 2001, 135, 125
\bibitem[Malkan, Gorjian \& Tam(1998)]{malkan98} Malkan, M. A., Gorjian, V. 
\& Tam, R. 1998, ApJ, 117, 25
\bibitem[Mihos \& Hernquist(1996)]{mihos96} Mihos, J. C., Hernquist, L. 1996,
ApJ, 464, 641
\bibitem[Mirabel \& Sanders(1989)]{mirabel89} Mirabel, I. F. \& Sanders, 
D. B. 1989, ApJ, 340, L53
\bibitem[Moshir et al.(1990)]{moshir90} Moshir, M., \& et al. 1990, IRAS Faint 
Source Catalogue, version 2.0
\bibitem[Murali et al.(2002)]{murali02}  Murali, C., Katz, N., Hernquist, 
L., Weinberg, D. H., \& Dave, R. 2002, ApJ, 571, 1
\bibitem[Pfefferkorn, Boller \& Rafanelli(2001)]{pfefferkorn01} 
Pfefferkorn, F., Boller, Th. \& Rafanelli, P. 2001, A\&A, 368, 797
\bibitem[Rafanelli \& Marziani(1992)]{rafanelli92} Rafanelli, P, \& 
Marziani, P 1992, AJ, 743, 1027 
\bibitem[Roberts(1969)]{roberts69} Roberts, M. S. 1969 AJ, 74, 859
\bibitem[Sakamoto et al.(1999)]{sakamoto99b} Sakamoto, K., Scoville, 
N. Z., Yun, M. S., Crosas, M., Genzel, R., \& Tacconi, L. J. 1999, ApJ, 514, 68
\bibitem[Sakamoto et al.(1999)]{sakamoto99} Sakamoto, K., Okumura, S. K.,
Ishizuki, S, Scoville, N. Z., 1999, ApJS, 124, 403 
\bibitem[Sanders \& Mirabel(1996)]{sanders96} Sanders, D. B., \& Mirabel, 
I. F., 1996, ARAA, 34, 749 
\bibitem[Sanders et al.(2003)]{sanders03} Sanders, D. B., 
Mazzarella, J. M., Kim, D. -C., Surace, J. A., \& Soifer, B. T. 
2003, AJ, 126, 1607
\bibitem[Scoville et al. (1993)]{Scoville93} Scoville, N. Z., Carlstrom, 
	J. E., Chandler, C. J., Phillips, J. A., Scott, S. L., Tilanus, 
	R. P. J., \& Wang, Z. 1993, PASP, 105, 1482
\bibitem[Scoville, Yun, \& Bryant(1997)]{sco97} Scoville, N. Z., Yun,
M. S., \& Bryant, P. M. 1997, ApJ, 484, 702
\bibitem[Sharp \& Keel(1985)]{sharp85} Sharp, N. A. \& Keel, W. C. 1985, 
AJ, 90, 469
\bibitem[Shepherd et al.(1994)]{shep94} Shepherd, M. C., Pearson, T. J.,
\& Taylor, G. B. 1994, BAAS, 26, 987
\bibitem[Soifer et al(1987)]{soifer87}  Soifer, B. T., Sanders, D. B., 
Madore, B. F., Neugebauer, G., Danielson, G. E., Elias, J. H., Lonsdale, 
Carol J. \& Rice, W. L. 1987, ApJ, 320, 238
\bibitem[Soifer et al(1989)]{soifer89} Soifer, B. T, Boehmer, L., 
Neugebauer, G. \& Sanders, D. B. 1989, AJ, 98, 766
\bibitem[Solomon \& Sage(1988)]{solomon88} Solomon, P. M. \& Sage, L. J. 
1988, ApJ, 334, 613
\bibitem[Struck(1999)]{struck99} Struck, C. 1999, PhR, 321, 1
\bibitem[Toomre \& Toomre(1972)]{toomre72} Toomre, A. \& Toomre, J. 1972, 
ApJ, 178, 623
\bibitem[Toomre(1977)]{toomre77} Toomre, A. 1977, in ``The Evolution of 
Galaxies and Stellar Populations,'' ed. B. M. Tinsley \& R. B. Larson 
(New Haven: Yale Univ.), p401
\bibitem[Tran et al.(2003)]{tran03} Tran, H. D., Sirianni, M., Ford,
H. C., Illingworth, G. D., Clampin, M. et al. 2003, ApJ, 585, 750
\bibitem[Trung et al.(2001)]{trung01} Trung, Dinh-V, Lo, K. Y., Kim, D.-C.,
Gao, Y. \& Gruendl, R. A. 2001, ApJ, 556, 141
\bibitem[Veilleux et al.(1995)]{veilleux95} Veilleux, S., Kim, D. -C., 
Sanders, D. B., Mazzarella, J. M. \& Soifer, B. T. 1995, ApJS, 98, 171
\bibitem[Veron, Goncalves \& Veron-Cetty(1997)]{veron97} Veron, P., 
Goncalaves, A. C., \& Veron-Cetty, M. -P. 1997, A\&A, 319, 52
\bibitem[Vorontsov-Velyaminov(1959)]{vv59} Vorontsov-Velyaminov,
B.A. 1959, Atlas and Catalogue of interacting galaxies. Part 1. 
Moscow University. Moscow
\bibitem[Vorontsov-Velyaminov(1977)]{vv77} Vorontsov-Velyaminov B.A.
1977, Atlas
 and Catalogue of interacting galaxies. Part 2. Moscow University. Moscow
\bibitem[Wilner \& Welch(1994)]{wilner94} Wilner, D. \& Welch, W. J. 
1994, ApJ, 427, 898
\bibitem[Xilouris et al.(2004)]{xilouris04} Xilouris, E. M., 
Georgakakis, A. E., Misiriotis, A. \& Charmandaris, V. 2004, MNRAS, 355, 57
\bibitem[Xu et al.(2000)]{xu00} Xu, C., Gao, Y., Mazzarella, J., Lu, N., 
Sulentic, J. W., Domingue, D. L. 2000, ApJ, 541, 644
\bibitem[Yao et al.(2003)]{yao03} Yao, L., Seaquist, E. R., Kuno, N. \& 
Dunne, L. 2003, ApJ, 588, 771
\bibitem[Young et al.(1986)]{young86}   Young, J. S., Kenney, J. D., 
Tacconi, L., Claussen, M. J., Huang, Y.-L., Tacconi-Garman, L., Xie, S. 
\& Schloerb, F. P. 1986, ApJ, 311, 17
\bibitem[Young et al.(1989)]{young89} Young, J. S.,  Xie, S, Kenney, 
J. D. P. \& Rice, W. L. 1989, ApJS, 70, 699
\bibitem[Young \& Scoville(1991)]{young91} Young, J. \& Scoville, N. Z.
1991, ARAA, 29, 581
\bibitem[Young et al.(1995)]{young95} Young, J. et al. 1995, ApJS, 98, 219
\bibitem[Young et al.(1996)]{young96} Young, J. S., Allen, L., Kenney, 
J. D. P., Lesser, A. \& Rownd, B. 1996, AJ, 112, 1903
\bibitem[Yun(1999)]{yun99} Yun, M. S. 1999, IAUS 186, 81
\bibitem[Yun \& Hibbard(2001)]{yun01} Yun, M. S., \& Hibbard, J. E. 2001, 
ApJ, 550, 104
\bibitem[Yun, Reddy \& Condon(2001)]{yun01b} Yun, M. S., Reddy, N. A. \& 
Condon, J. J. 2001, ApJ, 554, 803
\bibitem[Zhu et al.(1999)]{zhu99} Zhu, M., Seaquist, E. R., Davoust, E.,
Frayer, D. T. \& Bushouse, H. A.  1999, AJ, 118, 145 

\end{thebibliography}
\end{document}